\pdfoutput=1
\documentclass[10pt,a4paper]{article}
\usepackage[T2A]{fontenc}
\usepackage[utf8]{inputenc}
\usepackage[english]{babel}
\usepackage{amssymb,graphicx}
\usepackage{amsmath,amsfonts}
\usepackage{amsthm}
\usepackage{framed}
\usepackage{xcolor}
\usepackage{fullpage}
\usepackage[makeroom]{cancel}
\usepackage[export]{adjustbox} % Adjust the vertical alignment of the
\usepackage{microtype}
\usepackage{hyperref}

\title{\textbf{Mixed network calculus}}

\author{Yegor Zenkevich\thanks{yegor.zenkevich@gmail.com}\\
  {\small\textit{SISSA, via
      Bonomea 265, 34136 Trieste, Italy,}}\\
  {\small\textit{INFN, Sezione di Trieste,}}\\
  {\small\textit{IGAP, via Beirut 2/1, 34151 Trieste, Italy,}}\\
  {\small\textit{ITEP, Bolshaya Cheremushkinskaya street 25, 117218
      Moscow, Russia,}}\\
  {\small\textit{ITMP MSU, Leninskie gory 1, 119991 Moscow, Russia,}}\\
  {\small\textit{MIPT, Institutskii pereulok 9, 141700, Dolgoprudny, Russia}}}
\date{}
\begin{document}
\maketitle
\vspace{-46ex}
\noindent
{\textit{To T.} \hfill ITEP/TH-35/20}

{\hfill MIPT/TH-20/20}

\vspace{40ex}

\begin{abstract}
  We show how to combine higgsed topological vertices introduced in
  \cite{Zenkevich:2018fzl} with conventional refined topological
  vertices. We demonstrate that the extended formalism describes very
  general interacting D5-NS5-D3 brane systems. In particular, we
  introduce new types of intertwining operators of Ding-Iohara-Miki
  algebra between different types of Fock representations
  corresponding to the crossings of NS5 and D5 branes. As a byproduct
  we obtain an algebraic description of the Hanany-Witten brane
  creation effect, give an efficient recipe to compute the brane
  factors in $3d$ $\mathcal{N}=2$ and $\mathcal{N}=4$ quiver gauge
  theories and demonstrate how $3d$ $S$-duality appears in our setup.
\end{abstract}

\section{Introduction}
\label{sec:introduction}
Branes and algebras are related. Very generally, if a brane has some
other branes bound to it, then usually these states form a
representation of a ``brane algebra''. Probably the most well-known of
such setups is the Nakajima construction of the action of an affine
Lie algebra on the moduli space of instantons on
$\mathbb{R}^4$~\cite{Nakajima}. The development of the same ideas lead
to the AGT relation~\cite{Alday:2009aq}. In this paper we will
consider one example of a brane algebra, namely the Ding-Iohara-Miki
(DIM) algebra~\cite{DIM} and show that it describes surprisingly many
brane pictures. We will mostly focus on $3d$ theories living on D3
branes, but many different objects will also appear naturally in our
treatment.

The original motivation for this
investigation~\cite{Zenkevich:2018fzl} was to reinterpret the brane
constructions and dualities of $3d$ gauge theories with
$\mathcal{N}=2$ and $\mathcal{N}=4$ supersymmetry using the
``algebraic engineering'' approach. By this term we mean a dictionary
between brane pictures\footnote{Through a chain of dualities the same
  pictures can sometimes be interpreted as purely geometric
  backgrounds, e.g.\ toric diagrams of toric Calabi-Yau three-folds.}
and networks of intertwiners of certain typically infinite-dimensional
algebras. In particular, DIM algebra is a quantum toroidal algebra
$U_{q,t}(\widehat{\widehat{\mathfrak{gl}}}_1)$)\footnote{A
  generalization of the construction to
  $U_{q,t}(\widehat{\widehat{\mathfrak{gl}}}_n)$ is also
  available~\cite{Zenkevich:2019ayk}}, so it is ``doubly infinite'':
the generators span an integer plane $\mathbb{Z}^2$. Related
constructions for other algebras have been considered
in~\cite{Kimura:2015rgi}, and a different perspective on them was
proposed in~\cite{Gaiotto:2017euk}. In what follows we will frequently
use the notations and conventions for DIM algebra and its
representations taken from Appendix A of~\cite{Zenkevich:2018fzl}
without referring to particular sections. So by default one should
look up in there all the objects related DIM algebra used below.

The algebraic engineering approach that we will pursue here originated
from the interpretation~\cite{AFS} of the refined topological
vertex~\cite{Awata:2005fa, Iqbal:2007ii} as an intertwining operator
of the DIM algebra. In this approach the lines on the brane pictures
correspond to representations of the algebra while the intertwining
operators between representations play the role of brane
junctions. One draws Fock representations of the DIM algebra as
(solid) lines of rational slopes lying in a $2d$ plane:
\begin{equation}
  \label{eq:6}
  \includegraphics[valign=c]{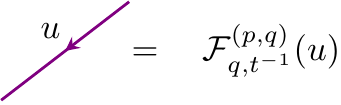}
\end{equation}
where $u$ denotes the spectral parameter. In the brane language these
lines can be understood as $(p,q)$ 5-branes\footnote{There is an
  unpleasant clash in our notation: $q$ denotes one of the charges of
  the 5-brane and also the $\Omega$-background deformation
  parameter. However, both conventions are so entrenched in the
  literature that we do not dare to change them and can only hope that
  there are not too many opportunities for confusion.}  of Type IIB
string theory forming a brane web. Their slope in the picture is
determined by their $(p,q)$-charges as follows from supersymmetry
constraints. The $SL(2,\mathbb{Z})$ automorphism group of the DIM
algebra acts on the plane of the picture and correspondingly on the
5-brane $(p,q)$-charges. It is therefore identified with the
$S$-duality group of the Type IIB string theory.

The triple junction of 5-branes corresponds to the intertwining
operators $\Psi_{q,t^{-1}} : \mathcal{F}^{(0,1)}_{q,t^{-1}}(w) \otimes
\mathcal{F}^{(1,0)}_{q,t^{-1}}(u)\to
\mathcal{F}^{(1,1)}_{q,t^{-1}}(-uw)$:
\begin{equation}
  \label{eq:1}
\Psi_{q,t^{-1}} \quad =  \quad \includegraphics[valign=c]{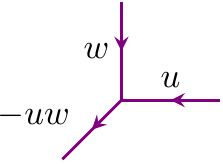}
\end{equation}
There is also the dual intertwiner $\Psi^{*}$ acting from a single
Fock representation to a pair of Fock representations. The explicit
expressions for the intertwiners $\Psi$ and $\Psi^{*}$ are given
in~\cite{AFS}.

The charges of the 5-branes have to be conserved at the junction so a
$(r_1,s_1)$ and $(r_2,s_2)$ branes join into a $(r_1+r_2,s_1+s_2)$
brane as drawn in the example above. The same is true for the slopes
of the branes, since they are related to the charges by
supersymmetry. The two $(p,q)$ charges correspond to the values of two
central charges of the Fock representation, e.g.\ the central charges
of $\mathcal{F}^{(r,s)}_{q,t^{-1}}$ are $\left( ( t/q )^{\frac{r}{2}},
  ( t/q )^{\frac{s}{2}}\right)$. The two central charges are denoted
by $(\gamma, \sqrt{\psi^{+}_0/\psi^{-}_0})$ in Appendix A
    of~\cite{Zenkevich:2018fzl}. See~\cite{Mironov:2016yue,
      Bourgine:2016vsq} for some applications of the algebraic
    formalism to 5-brane networks.

The deformation parameters $q$ and $t^{-1}$ of the DIM algebra are
interpreted as $\Omega$-background parameters of the theory. To
incorporate the permutation symmetry between $q$, $t^{-1}$ and $t/q$
one has to assume that in addition to the $\mathbb{R}^2$ plane of the
picture there are three more complex dimensions in the background. We
also know that the Fock representation
$\mathcal{F}^{(0,1)}_{q,t^{-1}}$ is the representation of the algebra
on the equivariant $K$-theory of the Hilbert scheme of points on
$\mathbb{C}^2$, which corresponds to a $5d$ $\mathcal{N}=1$ gauge
theory living on $\mathbb{C}_q \times \mathbb{C}_{t^{-1}} \times S^1$,
so the background should also contain an $S^1$. Thus, Type IIB theory
lives on $\mathbb{C}_q \times \mathbb{C}_{t^{-1}} \times
\mathbb{C}_{t/q} \times \mathbb{R}^2_{\mathrm{picture}} \times
\mathbb{R}\times S^1 $. Using the mnemonic from Appendix A.5 of~\cite{Zenkevich:2018fzl} the representations of the DIM algebra can
then be assigned to branes living in this background and fixed by the
equivariant $U(1)^2$ action on $\mathbb{C}^3$. The dictionary is
summarized in the table below\footnote{For example changing $\mathbb{C}_q \times
  \mathbb{C}_{t^{-1}}\times \mathbb{C}_{t/q}$ to
  $\widetilde{(\mathbb{C}^2/\mathbb{Z}_N})_{q,t^{-1}} \times
  \mathbb{C}_{t/q}$ corresponds to considering the algebra
  $U_{q,t}(\widehat{\widehat{\mathfrak{gl}}}_N)$ instead of
  $U_{q,t}(\widehat{\widehat{\mathfrak{gl}}}_1)$.}:
\begin{equation}
  \label{eq:7}
  \begin{array}{l|l|ccc|cc|c|c}
&&&&&\multicolumn{2}{c|}{\text{picture}}&t&\\
    \text{Brane}& \text{DIM rep} &\mathbb{C}_q&\mathbb{C}_{t^{-1}}&\mathbb{C}_{t/q}&\mathbb{R}&\mathbb{R}&\mathbb{R}_t&S^1\\
    \hline
    \mathrm{F1} & \text{generators} &&&&-&&&-\\
    \mathrm{D1} &\text{generators} &&&&&-&&-\\
    \hline
    \textcolor{violet}{\mathrm{D5}_{q,t^{-1}}} &
    \textcolor{violet}{\mathcal{F}^{(0,1)}_{q,t^{-1}}}&--&--&&&-&&-\\
    \textcolor{red}{\mathrm{D5}_{q,t/q}} &
    \textcolor{red}{\mathcal{F}^{(0,1)}_{q,t/q}}&--&&--&&-&&-\\
    \textcolor{blue}{\mathrm{D5}_{t^{-1},t/q}} &
    \textcolor{blue}{\mathcal{F}^{(0,1)}_{t^{-1},t/q}}&&--&--&&-&&-\\
    \hline
    \textcolor{violet}{\mathrm{NS5}_{q,t^{-1}}} &
    \textcolor{violet}{\mathcal{F}^{(1,0)}_{q,t^{-1}}}&--&--&&-&&&-\\
    \textcolor{red}{\mathrm{NS5}_{q,t/q}} &
    \textcolor{red}{\mathcal{F}^{(1,0)}_{q,t/q}}&--&&--&-&&&-\\
    \textcolor{blue}{\mathrm{NS5}_{t^{-1},t/q}} &
    \textcolor{blue}{\mathcal{F}^{(1,0)}_{t^{-1},t/q}}&&--&--&-&&&-\\
    \hline
    \textcolor{blue}{\mathrm{D3}_q} &
     \textcolor{blue}{\mathcal{V}_q}&--&&&&&-&-\\
     \textcolor{red}{\mathrm{D3}_{t^{-1}}} &
     \textcolor{red}{\mathcal{V}_{t^{-1}}}&&--&&&&-&-\\
     \textcolor{violet}{\mathrm{D3}_{t/q}} &
     \textcolor{violet}{\mathcal{V}_{t/q}}&&&--&&&-&-\\
     \hline
     \text{D7-$\overline{\mathrm{D7}}$ ?} &
     \mathcal{M}^{(0,c)}&--&--&--&&&-&-
\end{array}
\end{equation}

The first two lines in the table~\eqref{eq:7} are the examples of
$(p,q)$-strings, namely the fundamental string and the D1 brane. These
will play the role of the generators of the algebra. More precisely
DIM algebra generators will live at points where $(p,q)$ strings
attach to other branes. We will comment on this in
sec.~\ref{sec:d1-joins-d3}.

As we have already seen, 5-branes spanning different directions are
associated with different types of Fock representations. They are
depicted by lines (and entries in~\eqref{eq:7}) of different colors as
in~\cite{Zenkevich:2018fzl}: $(q,t^{-1})$ 5-branes are violet,
$(q,t/q)$ ones are red and $(t^{-1},t/q)$ are blue. Notice that the
triple junction~\eqref{eq:1} is allowed only when all three 5-branes
are of the same color. The spectral parameter $u$ of the Fock
representation $\mathcal{F}^{(r,s)}(u)$ corresponds to the position of
the brane in the plane of the picture (complexified by adding the
Wilson line along $S^1$ of the gauge field on the brane worldvolume).

The additional ingredient introduced in~\cite{Zenkevich:2018fzl} were
the D3 branes corresponding to three types of the so-called vector
representations $\mathcal{V}_q$, $\mathcal{V}_{t^{-1}}$ and
$\mathcal{V}_{t/q}$. Again, the branes are assigned colors according
to complex planes they span. The representation $\mathcal{V}_q$ of DIM
algebra is a representation by difference operators $x$ and $p =
q^{x \partial_x}$ on functions of a single
variable\footnote{In~\cite{Zenkevich:2018fzl} these representations
  had an additional spectral parameter, e.g.\
  $\mathcal{V}_q(w)$. However, a more natural way to understand $w$ is
  as a parameter of a \emph{state} inside the representation, the
  representation being independent of $w$.}. This representation is
invariant under the $SL(2,\mathbb{Z})$ automorphism group, as is the
D3 brane of Type IIB string theory.

Classically (with equivarant deformation turned off, $q\to 1$), the
variables $x$ and $p$ are (the complexifications of) the coordinates
in the plane of the picture $\mathbb{R}^2_{\mathrm{picture}}$ which
plays the role of the phase space. In the quantum regime the D3 branes
are delocalized in $\mathbb{R}^2_{\mathrm{picture}}$ and only their
wavefunctions, defined on the Lagrangian submanifolds are
meaningful. The dashed lines featuring in~\cite{Zenkevich:2018fzl} are
such wavefunctions with fixed horizontal position: a state in the
vector representation $\mathcal{V}_q$ with fixed $x = w$ is denoted by
a vertical dashed line with spectral parameter $w$:
\begin{equation}
\label{eq:9}
  \includegraphics[valign=c]{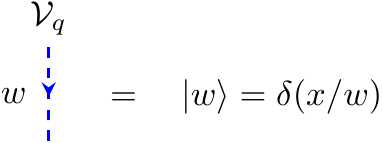}
\end{equation}

The higgsed vertices $\Phi$ introduced in~\cite{Zenkevich:2018fzl} are
the junctions between NS5 branes and D3 branes of various kinds. For
geometric reasons only junctions in which the NS5 and D3 branes share
a $\mathbb{C}$ plane are allowed. In our convention this means that
the colors of dashed and solid lines in the junction cannot coincide.
An example of a D3-NS5 junction is the operator $\Phi^q_{q,t^{-1}}:\mathcal{V}_q \otimes
\mathcal{F}^{(1,0)}_{q,t^{-1}}(u)\to \mathcal{F}^{(1,0)}_{q,t^{-1}}(t u)$:
\begin{equation}
  \label{eq:10}
 \Phi^q_{q,t^{-1}} (|w\rangle \otimes \ldots)  \quad = \quad  \includegraphics[valign=c]{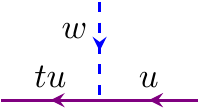}
\end{equation}
where we have already plugged in the state $|w\rangle \in
\mathcal{V}_q$ into the intertwiner. There is naturally, a dual
intertwiner $\Phi^{*q,t^{-1}}_q: \mathcal{F}^{(1,0)}_{q,t^{-1}}( u)
\to \mathcal{F}^{(1,0)}_{q,t^{-1}}(t^{-1}u) \otimes
\mathcal{V}_q$. One can find explicit expressions for the intertwiners
$\Phi$ and $\Phi^{*}$ in~\cite{Zenkevich:2018fzl}.

Let us briefly comment on the last row in the table~\eqref{eq:7}. Its
interpretation is not completely clear. The MacMahon representation
$\mathcal{M}^{(0,c)}$ of DIM algebra is the representation on plane
partitions (or $3d$ Young diagrams) and allows for an arbitrary value
of the central charge $c$. $\mathcal{M}^{(0,c)}$ can be built by
tensoring an infinite number of Fock spaces with specific spectral
parameters and then taking an irreducible part of the
product~\cite{semi-inf}. It turns out that for $c$ tuned to certain
quantized values only a \emph{finite} number of Fock spaces are needed
to build $\mathcal{M}^{(0,c)}$. It is therefore natural to assume that
this representation should be
\begin{enumerate}
\item related to the equivariant $K$-theory of the Hilbert scheme of
  points on $\mathbb{C}^3$,
\item for specific values of $c$ it should contain irreducible
  subrepresentations corresponding to stacks of 5-branes.
\end{enumerate}
This leads us to believe that the relevant system is a pair of D7 and
anti-D7 brane sitting $c$ distance apart in the plane of the
picture. However, such configuration seems to break all supersymmetry
making this interpretation doubtful. We will not focus on the MacMahon
representation here and only briefly mention it in
sec.~\ref{sec:cross-with-macm}.

In this short note we extend the dictionary proposed
in~\cite{Zenkevich:2018fzl} in several directions. In particular we
propose:
\begin{enumerate}
\item Intertwiners corresponding to junctions of D5 branes with D3
  branes. They correspond to pictures such as
\begin{equation}
  \label{eq:101}
  \tilde{\Phi}^{*q}_{q,t^{-1}}  \quad = \quad  \includegraphics[valign=c]{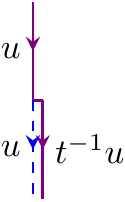}
\end{equation}
These operators can be described very explicitly using Young diagrams.

\item Brane crossings for a pair 5-branes and a 5-brane crossing
  3-brane. An example of a crossing between a D5 and an NS5 branes is
  \begin{equation}
    \label{eq:102}
    X^{q,t^{-1}}_{q,t^{-1}}  \quad = \quad  \includegraphics[valign=c]{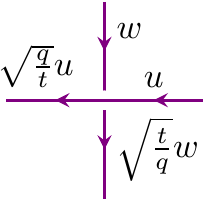}
  \end{equation}
  The crossings turn out to have a clear algebraic meaning: they are
  Lax matrices satisfying the Yang-Baxter relations with the DIM
  $R$-matrix~\cite{Awata:2016mxc,Awata:2016bdm,Awata:2018svb}.
  
\end{enumerate}

Employing these new ingredients we show how D3 branes arise from
5-brane configurations for quantized values of the spectral
parameters. We also clarify the relation between $S$-duality of
$\mathcal{N}=4$ $3d$ theories and the $SL(2,\mathbb{Z})$ automorphism
group of the quantum toroidal algebra
$U_{q,t}(\widehat{\widehat{\mathfrak{gl}}}_1)$. 

The plan of the paper is as follows. In sec.~\ref{sec:d3-brane-joins}
we introduce new types of intertwiners corresponding to a D3 brane
meeting a D5 brane. In sec.~\ref{sec:d1-joins-d3} we consider the
action of the algebra itself as an intertwining operator and related
it to $(p,q)$-string junctions. In sec.~\ref{sec:5-brane-crossings} we
introduce various brane crossings and clarify their algebraic
meaning. In sec.~\ref{sec:higgs-vert-from} we establish the precise
relation between the networks involving only the conventional refined
topological vertices and those with higgsed vertices. In
sec.~\ref{sec:tsun-ftsun-partition} using the new formalism we
consider the brane construction of $3d$ $T[SU(N)]$ theory and its
cousins and analyze their $S$-duality properties. We present our
conclusions in sec.~\ref{sec:conclusions}.

\section{D3 brane joins D5 brane}
\label{sec:d3-brane-joins}
In~\cite{Zenkevich:2018fzl} we have introduced the D3-NS5 junctions
such as~\eqref{eq:10}. Here we would like to add the D3-D5 junctions
to our vocabulary. In our pictorial notation these junctions are
between a (vertical) dashed line corresponding to vector
representation e.g.\ $\mathcal{V}_q$ meeting a vertical solid line
corresponding to a Fock representation. As in the case of D3-NS5
junctions, the merging branes should share $\mathbb{C}_q$ part of
their worldvolume so D3$_q$ brane can meet only D5$_{q,t^{-1}}$ and
D5$_{q,t/q}$ corresponding to the Fock representations
$\mathcal{F}^{(0,1)}_{q,t^{-1}}$ and $\mathcal{F}^{(0,1)}_{q,t/q}$
respectively. We denote the corresponding intertwiners by
$\tilde{\Phi}^{*q}_{q,t^{-1}}$ and $\tilde{\Phi}^{*q}_{q,t/q}$
respectively and draw them as follows:
\begin{equation}
  \label{eq:11}
   \tilde{\Phi}^{*q}_{q,t^{-1}}  \quad =
   \quad  \includegraphics[valign=c]{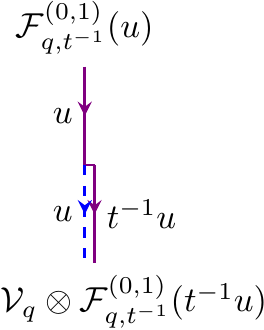} \qquad \qquad    \tilde{\Phi}^{*q}_{q,t/q}  \quad =
   \quad  \includegraphics[valign=c]{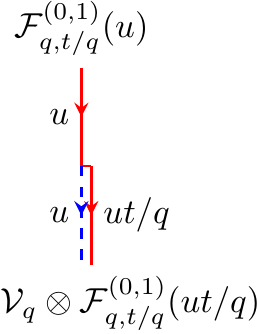}
\end{equation}
Notice the shift of the spectral parameter of the Fock representation
after the junction. This shift is the same as in the D3-NS5 junctions
$\Phi^{*q}_{q,t^{-1}}$ and $\Phi^{*q}_{q,t/q}$ mentioned after
Eq.~\eqref{eq:10}.

One can deduce the explicit expression for the D3-D5 intertwiners from
the semi-infinite construction of the vertical Fock representations of
DIM algebra~\cite{semi-inf}. The vertical Fock representation
$\mathcal{F}^{(0,1)}_{q,t^{-1}}(u)$ can be understood as a
(appropriately regularized) limit of an infinite tensor product of
certain subspaces of the vector representations $\mathcal{V}_q$. In
particular a state in the standard basis\footnote{This is usually
  taken to be the basis of Macodnald polynomials however this will not
  be important for us.} $|\lambda,u \rangle \in
\mathcal{F}^{(0,1)}_{q,t^{-1}}(u)$ parametrized by a Young diagram
$\lambda = [\lambda_1, \lambda_2, \ldots]$ is obtained from the
product
\begin{equation}
  \label{eq:12}
  |\lambda,u \rangle_{\mathcal{F}^{(0,1)}_{q,t^{-1}}(u)} =
  \sideset{}{'}\lim_{N \to \infty} | u q^{\lambda_1} \rangle_{\mathcal{V}_q} \otimes | u q^{\lambda_2}
  t^{-1} \rangle_{\mathcal{V}_q} \otimes | u q^{\lambda_3} t^{-2} \rangle_{\mathcal{V}_q} \otimes
  \ldots \otimes | u t^{-N+1} \rangle_{\mathcal{V}_q},
\end{equation}
where the prime denotes the regularization. Taking the first factor
from the tensor product and treating it as an independent vector
representation\footnote{This argument is similar to the Hilbert's
  Grand Hotel paradox: if a guest in the first room of a fully
  occupied infinite hotel moves out, the hotel manager can still
  retain full occupancy by moving every guest from room $i$ to room
  $i-1$.} one arrives at the formula for the intertwiner
\begin{equation}
  \label{eq:13}
\boxed{  \tilde{\Phi}^{*q}_{q,t^{-1}} |\lambda,
  u\rangle_{\mathcal{F}^{(0,1)}_{q,t^{-1}}(u)} = |q^{\lambda_1}
  u\rangle_{\mathcal{V}_q} \otimes | \lambda \backslash \lambda_1,
  t^{-1} u \rangle_{_{\mathcal{F}^{(0,1)}_{q,t^{-1}}(t^{-1}u)}},}
\end{equation}
where $\lambda \backslash \lambda_1$ denotes the Young diagram
$[\lambda_2, \lambda_3, \ldots]$. The formulas for 
$\tilde{\Phi}^{*q}_{q,t/q}$ and for all other allowed combinations of
Fock and vector spaces are obtained by permuting $(q,t^{-1},t/q)$.

Using the explicit formulas for the action of DIM algebra generators
on the vertical Fock (see e.g.~\cite{AFS}) and vector representation
(see Appendix A.4 of~\cite{Zenkevich:2018fzl}) one can directly verify
that the operator defined by~\eqref{eq:13} indeed satisfies the
intertwining relations
\begin{equation}
  \label{eq:14}
  \tilde{\Phi}^{*q}_{q,t^{-1}}g = \Delta(g)\tilde{\Phi}^{*q}_{q,t^{-1}},
\end{equation}
where $\Delta$ is the ``vertical'' DIM coproduct (see A.2
of~\cite{Zenkevich:2018fzl}). From the symmetry between
$(q,t^{-1},t/q)$ we find that all junctions of this type are given by
the formulas similar to~\eqref{eq:13}.

\section{$(p,q)$-string joins D3 brane: the brane nature of DIM
  representations}
\label{sec:d1-joins-d3}
In this section we briefly explain how representations of DIM algebra
arise from junctions of $(p,q)$ strings with a given brane. The type
of the brane naturally determines the representation of the algebra as
summarized in table~\eqref{eq:7}. We will focus of the simplest
example of the D3$_q$ brane corresponding to the vector representation
$\mathcal{V}_q$. The action of DIM algebra currents $x^{\pm}(z)$,
$\psi^{\pm}(z)$ in this representation are given by
 \begin{align}
   x^{+}(z)|w\rangle_{\mathcal{V}_q} &= - \frac{1}{1-q^{-1}} \delta
     \left( \frac{w}{z} \right) \left| q w
    \right\rangle_{\mathcal{V}_q},\label{eq:69} \\
    x^{-}(z)|w\rangle_{\mathcal{V}_q} &= - \frac{1}{1-q} \delta \left(
      \frac{w}{qz} \right) \left| \frac{w}{q}
    \right\rangle_{\mathcal{V}_q},\label{eq:70}
    \\
    \psi^{+}(z) |w\rangle_{\mathcal{V}_q} &= \frac{\left( 1 -
        \frac{t}{q} \frac{w}{z}\right) \left( 1 - \frac{1}{t}
        \frac{w}{z}\right)}{\left( 1 - \frac{w}{z} \right) \left( 1 -
        \frac{1}{q} \frac{w}{z} \right)}
    |w\rangle_{\mathcal{V}_q} ,\\
      \psi^{-}(z) |w\rangle_{\mathcal{V}_q} &= \frac{\left( 1 -
            \frac{q}{t} \frac{z}{w}\right) \left( 1 - t
            \frac{z}{w}\right)}{\left( 1 - \frac{z}{w} \right) \left(
            1 - q \frac{z}{w} \right)} |w\rangle_{\mathcal{V}_q}
          . \label{eq:68}
 \end{align}
 DIM currents can be expanded as series in $z$ with coefficients
 $P_{(r,s)}$:
 \begin{align}
   \label{eq:71}
   x^{\pm}(z) &= \sum_{n\in \mathbb{Z}} P_{(n,\pm 1)} q^{\frac{\pm
       1-n}{2}}z^{-n},\\
   \psi^{\pm}(z) &= \exp \left[ \sum_{n \geq 1} \frac{\kappa_n}{n}
     P_{(\pm n,0)} q^{\mp \frac{n}{2}} z^{\mp n} \right].
 \end{align}
 The generators $P_{(r,s)}$ corresponds to a junction of an
 $(r,s)$-string with the D3 brane. We draw $(r,s)$-string as a black
 wavy line in the plane of the picture at slope $(r,s)$:
 \begin{equation}
   \label{eq:72}
   \includegraphics[valign=c]{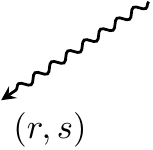} \quad = \quad
   \text{$(r,s)$-string.}
 \end{equation}
 The string is colorless for the following reason. Recall from
 table~\eqref{eq:7} that $(r,s)$-strings don't span any
 $\mathbb{C}$-planes on which the equivariant group acts, and
 therefore remain invariant under the permutation of
 $(q,t^{-1},t/q)$. Indeed, this coincides with what we know about the
 DIM algebra: it is symmetric under such permutations.

 The junction of a $(0,1)$-string with a D3$_q$ brane corresponds to
 the following relation following from~\eqref{eq:69}:
 \begin{equation}
   \label{eq:73}
   P_{(0,1)}|w\rangle_{\mathcal{V}_q} = -\frac{1}{q^{1/2}-q^{-1/2}}|qw\rangle_{\mathcal{V}_q},
 \end{equation}
 Indeed, using the wavy line Eq.~\eqref{eq:73} can be drawn as
 follows:
 \begin{equation}
   \label{eq:74}
   \includegraphics[valign=c]{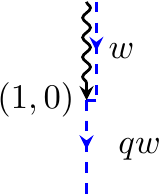}
 \end{equation}
 The junction changes the position of the D3 brane by $q$ in the same
 way as the D3 brane joining D5 brane changes its spectral parameter
 in Eq.~\eqref{eq:11}. Similarly we can draw the action of a
 $(0,-1)$-string:
 \begin{equation}
   \label{eq:67}
  P_{(0,-1)}|w\rangle_{\mathcal{V}_q} =
  -\frac{1}{q^{-1/2}-q^{1/2}}\left| \frac{w}{q} \right\rangle_{\mathcal{V}_q} \qquad \qquad \Leftrightarrow \qquad \qquad \includegraphics[valign=c]{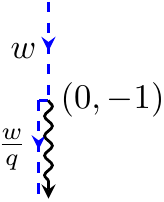}
 \end{equation}
 Fundamental strings don't change the position of the D3 brane:
 \begin{equation}
   \label{eq:75}
   P_{(1,0)}|w\rangle_{\mathcal{V}_q} =
  -\frac{w}{q^{-1/2}-q^{1/2}}|w\rangle_{\mathcal{V}_q} \qquad \qquad \Leftrightarrow \qquad \qquad \includegraphics[valign=c]{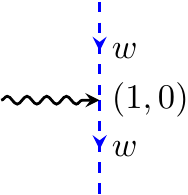}
 \end{equation}
 Similarly we could have drawn pictures for other more complicated
 representations associated with 5-branes. However, in that case the
 $(p,q)$-string junction would not change the position of the brane,
 but only alter the state (the Young diagram) living on it.

\section{5-brane crossings}
\label{sec:5-brane-crossings}
The next new element we are going to introduce is what we call the
5-brane crossings, though the branes really cross only in the
projection to the plane of the picture,
$\mathbb{R}^2_{\mathrm{picture}}$. In reality the operators we are
going to introduce correspond to the \emph{skew} branes: pairs of
lines approaching each other in a $3d$ space but failing to intersect.

In sec.~\ref{sec:introduction} we have summarized that 5-branes of the
same color from table~\eqref{eq:7} spanning lines in the plane of the
picture $\mathbb{R}^2_{\mathrm{picture}}$ can join into a brane web
and that this web corresponds to a network of intertwiners of Fock
representations of DIM algebra. However, in this construction we have
implicitly assumed that the 5-branes lie at the same value of $t$ on
the $\mathbb{R}_t$ line --- otherwise they have no chance to meet.

One can wonder what algebraic object corresponds to a D5$_{q,t^{-1}}$
and an NS5$_{q,t^{-1}}$ branes lying at \emph{different} values of
$t$. Two such branes lie along two skew lines in the $3d$ space
$\mathbb{R}^2_{\mathrm{picture}}\times \mathbb{R}_t$ never touching
each other, so one might think that the corresponding operator should
be trivial. This intuition turns out to be correct for $q=t$, however,
in general it fails and the brane crossing is a nontrivial operator on
the tensor product of two Fock spaces which we denote by
$X^{q,t^{-1}}_{q,t^{-1}}$. In fact there are two distinct situations:
either NS5 brane lies above D5 brane or below it (we imagine positive
$t$ direction goes upwards):
\begin{equation}
  \label{eq:15}
  X^{q,t^{-1}}_{q,t^{-1}}(u,w) \quad =\quad  \includegraphics[valign=c]{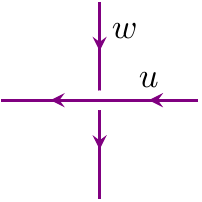} \qquad
  \qquad \qquad \bar{X}^{q,t^{-1}}_{q,t^{-1}}(u,w) \quad =\quad \includegraphics[valign=c]{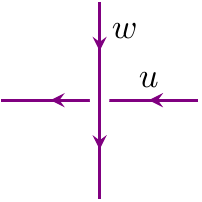}
\end{equation}

\subsection{5-brane crossing arise from degenerate resolved picture}
\label{sec:degen-resolv-conif}
Both of the situations depicted in~\eqref{eq:15} can be continuously deformed into the picture
where the two branes lie in the same plane of constant $t$ and do
touch each other. When this happens one can ``resolve'' the crossing
as follows:
\begin{equation}
  \label{eq:16}
  C(u,w,Q) \quad =\quad     \includegraphics[valign=c]{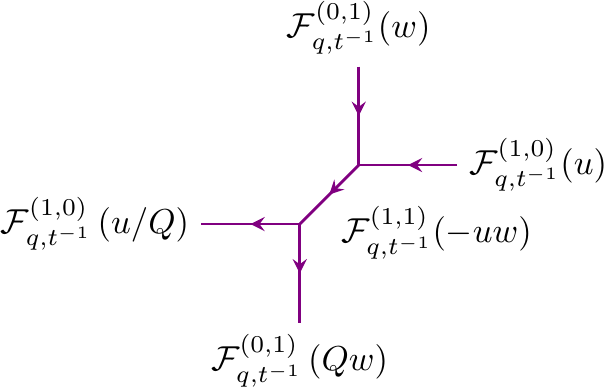}
\end{equation}
The resolution is parametrized by a complex number $Q$. The first
important assumption in our search for the operators
$X^{q,t^{-1}}_{q,t^{-1}}$ and $\bar{X}^{q,t^{-1}}_{q,t^{-1}}$
corresponding to~\eqref{eq:15} is that they are obtained from the
resolved crossing $C(u,w,Q)$ by specializing to certain particular
values of $Q$. In other words, we assume that when $Q$ is tuned to
these special values the branes are allowed to detach from each other
out of the constant $t$ slice of the $3d$ space and when this happens
the operators don't change any further. In particular, they are
\emph{independent} of the separation of the branes along
$\mathbb{R}_t$. They do, however, depend on the ordering along this
direction, so that the two operators $X^{q,t^{-1}}_{q,t^{-1}}$ and
$\bar{X}^{q,t^{-1}}_{q,t^{-1}}$ in~\eqref{eq:15} are distinct.

\subsection{5-brane crossing is the DIM Lax operator}
\label{sec:inters-5-bran}
The second key observation from which we deduce the crossing operators
is that when the branes are separated in the $t$ direction the
corresponding operators must satisfy the Yang-Baxter equation. We
already know from~\cite{Awata:2016bdm} that the operators $C(Q)$
satisfy the Yang-Baxter equation with the DIM $R$-matrix \emph{up to a
  function of spectral parameters,} which we call an anomaly. However,
for two special values of $Q_{\pm} = (q/t)^{\pm \frac{1}{2}}$ the
anomaly vanishes, so the operators $C \left( (q/t)^{\pm \frac{1}{2}}
\right)$ do satisfy the Yang-Baxter equation exactly. They are
therefore \emph{specializations} of the universal DIM $R$-matrix
$\mathcal{R}$ in a tensor product of a vertical and horizontal Fock
representations. Moreover, these two operators are inverses of each
other. We therefore postulate that
\begin{equation}
      \label{eq:18}
  \boxed{\begin{aligned}
  X^{q,t^{-1}}_{q,t^{-1}}(u,w) &= C \left(u,w, (t/q)^{\frac{1}{2}}
  \right) = \mathcal{R}|_{\mathcal{F}^{(0,1)}_{q,t^{-1}}(w) \otimes
    \mathcal{F}^{(0,1)}_{q,t^{-1}}(u)},\\
  \bar{X}^{q,t^{-1}}_{q,t^{-1}}(u,w) &= C \left(u,w, (q/t)^{\frac{1}{2}} \right)=
  \mathcal{R}^{-1}|_{\mathcal{F}^{(0,1)}_{q,t^{-1}}(w)
    \otimes \mathcal{F}^{(0,1)}_{q,t^{-1}}(u)},% \label{eq:19}
  \end{aligned}}
\end{equation}
In other words, the crossing operators are \emph{Lax operators} of DIM
algebra, satisfying the $RLL$ relations.

Let us give some explicit formulas for the crossing operator. We do
this in two different ways and demonstrate that the answers match:
\begin{enumerate}
\item In sec.~\ref{sec:lax-matrix-from} we combine explicit
  expressions for the 5-brane junctions $\Psi_{q,t^{-1}}$
  from Eq.~\eqref{eq:1} and $\Psi^{*}_{q,t^{-1}}$.
\item In sec.~\ref{sec:lax-matrix-from-1} we evaluate the universal DIM $R$-matrix $\mathcal{R}$ in the
  tensor product of the vertical and horizontal Fock representations.
\end{enumerate}

\subsubsection{Lax operator from two 5-brane junctions}
\label{sec:lax-matrix-from}
For simplicity we will limit ourselves to the matrix elements of the
crossing operators between the vacuum states in the vertical Fock
space, which we call $\chi_{q,t^{-1}}^{q,t^{-1}}(u,w)$ and
$\bar{\chi}_{q,t^{-1}}^{q,t^{-1}}(u,w)$:
\begin{align}
  \label{eq:30}
  \chi_{q,t^{-1}}^{q,t^{-1}}(u,w) &=
  (  \ldots \otimes \langle \varnothing, \sqrt{t/q} \, w | )
  X_{q,t^{-1}}^{q,t^{-1}}(u,w)(|\varnothing,w\rangle \otimes
  \ldots) \quad =\quad  \includegraphics[valign=c]{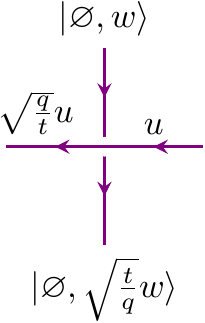},\\
    \bar{\chi}_{q,t^{-1}}^{q,t^{-1}}(u,w) &=
  (  \ldots \otimes \langle \varnothing, \sqrt{q/t} \, w | ) X_{q,t^{-1}}^{q,t^{-1}}(u,w)(|\varnothing,w\rangle \otimes \ldots)\quad =\quad  \includegraphics[valign=c]{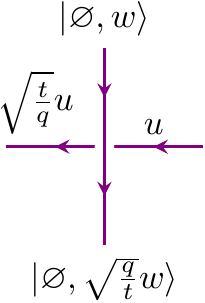}
\end{align}

$\chi_{q,t^{-1}}^{q,t^{-1}}(u,w)$ and
$\bar{\chi}_{q,t^{-1}}^{q,t^{-1}}(u,w)$ are vertex operators acting on
the horizontal Fock space. To calculate them we use explicit
expressions for the 5-brane junctions with vacuum state on one of the
legs~\cite{AFS}:
\begin{align}
  \label{eq:17}
  \Psi_{q,t^{-1}} (|\varnothing,w \rangle \otimes \ldots) &= (-w)^Q
  :\exp \left[\sum_{n \neq 0} \frac{w^{-n}}{n}\frac{a_n}{1-q^{-n}}
  \right]:,\\
 ( \ldots \otimes \langle \varnothing, (t/q)^{\pm \frac{1}{2}} w |) \Psi^{*}_{q,t^{-1}}  &= (-w)^{-Q}
  :\exp \left[- \sum_{n \neq 0} \frac{w^{-n}}{n}\left( \frac{t}{q}
    \right)^{\frac{|n| \mp n}{2}}  \frac{a_n}{1-q^{-n}}
  \right]:,
\end{align}
where $a_n$ form a Heisenberg algebra acting on the horizontal Fock
space:
\begin{equation}
  \label{eq:32}
  [a_n, a_m] = n \frac{1 - q^{|n|}}{1-t^{|n|}} \delta_{n+m,0},
\end{equation}
and $Q$ is the operator shifting the spectral parameter of the Fock
space:
\begin{equation}
  \label{eq:33}
  (-w)^Q | u, \lambda\rangle = |
  -w u, \lambda\rangle.
\end{equation}
Combining the intertwiners as in the picture~\eqref{eq:16} we get the
expressions for the crossing operators:
\begin{equation}
  \label{eq:34}
  \boxed{
\begin{aligned}
  \chi_{q,t^{-1}}^{q,t^{-1}}(u,w) &= (q;q,t)_{\infty} \left(
    \frac{q}{t} \right)^{\frac{Q}{2}} \exp \left[ -\sum_{n \geq 1}
    \frac{w^n}{n} \frac{\left( 1- (t/q)^n \right)}{1-q^n} a_{-n} \right],\\
  \bar{\chi}_{q,t^{-1}}^{q,t^{-1}}(u,w) &= (t;q,t)_{\infty} \left(
    \frac{t}{q} \right)^{\frac{Q}{2}} \exp \left[ \sum_{n \geq 1}
    \frac{w^{-n}}{n} \frac{\left( 1- (t/q)^n \right)}{1-q^{-n}} a_n \right],% \label{eq:37}
\end{aligned}}
\end{equation}
where
\begin{equation}
  \label{eq:35}
  (x;q,t)_{\infty} = \prod_{i,j \geq 0} (1-q^i t^j x).
\end{equation}
The numerical prefactors in Eqs.~\eqref{eq:34} are
inessential and we drop them henceforth.

\subsubsection{Lax operator from the universal DIM $R$-matrix}
\label{sec:lax-matrix-from-1}
First of all, we observe that there are shifts in spectral parameters
of both lines participating the crossing:
\begin{equation}
  \label{eq:20}
    X^{q,t^{-1}}_{q,t^{-1}}(u,w) \quad =\quad  \includegraphics[valign=c]{figures/d5-ns5-simpl-crop} \qquad
  \qquad \qquad \bar{X}^{q,t^{-1}}_{q,t^{-1}}(u,w) \quad =\quad \includegraphics[valign=c]{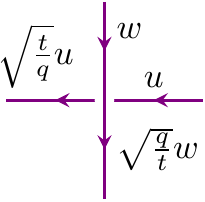}
\end{equation}
These shifts match with the structure of the universal $R$-matrix,
which can be written as a product of four parts (see~\cite{Negut-R, FJMM-Bethe}):
\begin{equation}
  \label{eq:21}
  \mathcal{R} = P \mathcal{R}_0 \mathcal{R}_1 \mathcal{R}_2,
\end{equation}
where $P$ is the permutation operator for the two tensor factors and
\begin{equation}
  \label{eq:22}
  \mathcal{R}_0 = e^{-c \otimes d - d \otimes c - c_{\perp} \otimes d_{\perp} - d_{\perp} \otimes c_{\perp}},
\end{equation}
where $c$, $c_{\perp}$ and $d$, $d_{\perp}$ are the central charges
and the grading operators of the DIM algebra respectively. The
$\mathcal{R}_1$ and $\mathcal{R}_2$ parts of the $R$-matrix have
Khoroshkin-Tolstoy form\footnote{The sign in the exponent in the
  $\mathcal{R}_1$ part is different in~\cite{FJMM-Bethe}
  and~\cite{FJMM-Bethe-2}. We use that of~\cite{FJMM-Bethe}. We also
  use the modified definitions of $H^{\pm}_n$ compared
  to~\cite{FJMM-Bethe, FJMM-Bethe-2}: $(H^{\pm}_n)_{\mathrm{our}} =
  \gamma^{\frac{n}{2}} (H^{\pm}_n)_{\mathrm{\cite{FJMM-Bethe,
        FJMM-Bethe-2}}}$, corresponding to
  $(\psi^{\pm}(z))_{\mathrm{our}} = (\psi^{\pm}(\gamma^{-1/2} z))_{\mathrm{\cite{FJMM-Bethe,
        FJMM-Bethe-2}}}$.}~\cite{KhorTolst}:
\begin{gather}
  \label{eq:27}
  \mathcal{R}_1 = \exp \left[ -\sum_{n \geq 1}
    \frac{n}{\kappa_n} \gamma^{-\frac{n}{2}} H^{+}_n \otimes
      \gamma^{\frac{n}{2}} H^{-}_{-n}\right]\\
    \mathcal{R}_2= \prod^{\curvearrowleft}_{s = \frac{a}{b} \in
      (0,\infty)} \mathcal{R}_s = 1 + \kappa_1 \sum_{i\in \mathbb{Z}}
    x^{+}_i \otimes x^{-}_{-i} + \ldots\label{eq:28}
\end{gather}
where $\kappa_n = (1-q^n)(1-t^{-n})(1-(t/q)^n)$, the definitions of
$x^{\pm}_i$ are given in Appendix A of~\cite{Zenkevich:2018fzl} and
the generators $H^{\pm}_n$ are related to $\psi^{\pm}(z)$
of~\cite{Zenkevich:2018fzl} by
\begin{equation}
  \label{eq:3}
  \psi^{\pm}(z) = \psi^{\pm}_0 \exp \left[ \sum_{n \geq 1}
    H^{\pm}_{\pm n} z^{\mp n} \right].
\end{equation}
The element $\mathcal{R}_2$ is given by an infinite product of terms,
whose explicit form is given e.g.\ in~\cite{Negut-R}. What will be
important for us is that there is that $\mathcal{R}_2$ can be written
as a graded expansion in the grading counted by $d_{\perp}$. This
expansion starts as in Eq.~\eqref{eq:28} with $x^{+}_i$ (resp.\
  $x^{-}_{-i}$) having grading $1$ (resp.\ $(-1)$).
  
The part which produces the shifts of the spectral parameters
in~\eqref{eq:20} is $\mathcal{R}_0$. To see this we notice that the
central charges of the horizontal Fock representation are
\begin{equation}
  \label{eq:23}
  (e^c, e^{c_{\perp}})|_{\mathcal{F}^{(1,0)}_{q,t^{-1}}(u)} = \left(
    \sqrt{\frac{t}{q}}, 1 \right),
\end{equation}
while the vertical representation is obtained by acting with
$SL(2,\mathbb{Z})$ transformation $\left(
\begin{smallmatrix}
  0& 1\\
  -1& 0
\end{smallmatrix}
\right)$ on Eq.~\eqref{eq:23}:
\begin{equation}
  \label{eq:24}
  (e^c, e^{c_{\perp}})|_{\mathcal{F}^{(0,1)}_{q,t^{-1}}(u)} = \left(
    1, \sqrt{\frac{q}{t}} \right).
\end{equation}
The grading operator $d$ acts on the horizontal Fock
representation as a box-counting operator:
\begin{equation}
  \label{eq:25}
  d | \lambda,u\rangle_{\mathcal{F}^{(1,0)}_{q,t^{-1}}(u)} = |\lambda| | \lambda,u\rangle_{\mathcal{F}^{(1,0)}_{q,t^{-1}}(u)},
\end{equation}
while $d_{\perp}$ is strictly speaking not defined as an operator on
the horizontal Fock space. It shifts the spectral parameter $u$ of the
representation:
\begin{equation}
  \label{eq:26}
    Q^{ d_{\perp}} |
    \lambda,u\rangle_{\mathcal{F}^{(1,0)}_{q,t^{-1}}(u)} =  | \lambda,
    Q u\rangle_{\mathcal{F}^{(1,0)}_{q,t^{-1}}(Q u)}.
\end{equation}
Therefore, the operator $e^{-c_{\perp} \otimes d_{\perp}}$ acts by
shifting the spectral parameter of the \emph{horizontal} Fock
representation by the value of the central charge of the
\emph{vertical} Fock representation. Combining all grading operators
together we get
\begin{equation}
  \label{eq:29}
  \mathcal{R}_0 |\mu,w\rangle_{\mathcal{F}^{(0,1)}_{q,t^{-1}}(w)} \otimes |\lambda,u\rangle_{\mathcal{F}^{(1,0)}_{q,t^{-1}}(u)} = \left(
    \frac{t}{q} \right)^{\frac{|\lambda|-|\mu|}{2}}   \left| \mu,
    \sqrt{\frac{t}{q}} w
  \right\rangle_{\mathcal{F}^{(0,1)}_{q,t^{-1}}\left(\sqrt{\frac{t}{q}} w\right)} \otimes
  \left|\sqrt{\frac{q}{t}} \lambda,u
  \right\rangle_{\mathcal{F}^{(1,0)}_{q,t^{-1}}\left(\sqrt{\frac{q}{t}}
      u\right)},
\end{equation}
This is the pattern of the spectral parameter shifts drawn
in~\eqref{eq:20}. It can also be written using the momentum shift
operator $Q$ as in Eq.~\eqref{eq:34}.

As in sec.~\ref{sec:lax-matrix-from} we are interested in vacuum
matrix elements~\eqref{eq:30} of the Lax operator. These matrix
elements can be calculated explicitly from the universal $R$-matrix
since the part $\mathcal{R}_2$ of the $R$-matrix~\eqref{eq:21} does
not contribute. This happens because all elements of negative
$d_{\perp}$ grading annihilate the vertical ket vacuum
$|\varnothing,w\rangle$ and those of positive grading annihilate the
bra vacuum $\langle \varnothing, (q/t)^{\pm 1/2} w|$.

The remaining $\mathcal{R}_1$ part of the $R$-matrix is also easy to
calculate if we recall the expressions for the DIM algebra generators
$H_n^{\pm}$ for vertical and horizontal Fock representations (see
e.g.~\cite{AFS}):
\begin{align}
  \label{eq:31}
  H^{+}_n|\varnothing, w \rangle_{\mathcal{F}^{(0,1)}_{q,t^{-1}}(w)} &=
    \frac{w^n}{n} \left( 1 - (t/q)^n \right)|\varnothing,w
    \rangle_{\mathcal{F}^{(0,1)}_{q,t^{-1}}(w)},\\
    H^{-}_{-n}|\varnothing,w
    \rangle_{\mathcal{F}^{(0,1)}_{q,t^{-1}}(w)} &= \frac{w^{-n}}{n}
    \left( 1 - (q/t)^n \right)|\varnothing,w
    \rangle_{\mathcal{F}^{(0,1)}_{q,t^{-1}}(w)}.\\
    H^{+}_n|_{\mathcal{F}^{(1,0)}_{q,t^{-1}}(u)}
      &= - \frac{(1-t^n) \left( 1 - \left( \frac{t}{q} \right)^n
        \right)}{n} \left( \frac{q}{t} \right)^{\frac{n}{4}}
      a_n,\\
      H^{-}_{-n}|_{\mathcal{F}^{(1,0)}_{q,t^{-1}}(u)} &= \frac{(1-t^{-n})
        \left( 1 - \left( \frac{t}{q} \right)^n \right)}{n} \left( \frac{q}{t} \right)^{\frac{n}{4}}
      a_{-n},\label{eq:56}
\end{align}
where $a_n$ are the Heisenberg generators
satisfying~\eqref{eq:32}. Therefore
\begin{equation}
  \label{eq:36}
 \mathcal{R}_1
  (|\varnothing,w\rangle_{\mathcal{F}^{(0,1)}_{q,t^{-1}}(w)} \otimes
  \ldots) = \exp \left[ -\sum_{n \geq 1} \frac{w^n}{n} \frac{\left(1-
        (t/q)^n
      \right)}{(1-q^n)} a_{-n} \right],
\end{equation}
and
\begin{equation}
  \label{eq:54}
  \chi_{q,t^{-1}}^{q,t^{-1}}(u,w) = \left(
    \frac{q}{t} \right)^{\frac{Q}{2}} \exp \left[ -\sum_{n \geq 1}
    \frac{w^n}{n} \frac{\left( 1- (t/q)^n \right)}{1-q^n} a_{-n} \right]
\end{equation}
exactly as in Eq.~\eqref{eq:34}.

To get $\bar{\chi}_{q,t^{-1}}^{q,t^{-1}}(u,w)$ we should compute the
inverse of the universal $R$-matrix, for which extra caution is needed
because the factors $\mathcal{R}_0$, $\mathcal{R}_1$ and
$\mathcal{R}_2$ do not commute with each other. We begin by noting
that
\begin{equation}
  \label{eq:50}
  \mathcal{R}^{-1} = \mathcal{R}_2^{-1} \mathcal{R}_1^{-1}
  \mathcal{R}_0^{-1} P = P (\mathcal{R}_2^{\mathrm{op}})^{-1} (\mathcal{R}_1^{\mathrm{op}})^{-1}
  (\mathcal{R}_0^{\mathrm{op}})^{-1},
\end{equation}
where $^{\mathrm{op}}$ denotes the exchange of two factors in the
tensor product. The parts which will be relevant for us are
$\mathcal{R}_0$ and $\mathcal{R}_1$, for which we have
\begin{align}
  \label{eq:52}
    (\mathcal{R}_0^{\mathrm{op}})^{-1}&= e^{c \otimes d + d \otimes c + c_{\perp} \otimes d_{\perp} + d_{\perp} \otimes c_{\perp}},\\
  (\mathcal{R}_1^{\mathrm{op}})^{-1} &= \exp \left[ \sum_{n \geq 1}
    \frac{n}{\kappa_n}  
      \gamma^{\frac{n}{2}} H^{-}_{-n}\otimes \gamma^{-\frac{n}{2}} H^{+}_n\right].
\end{align}
We then would like to move $(\mathcal{R}^{\mathrm{op}}_0)^{-1}$ past
$(\mathcal{R}^{\mathrm{op}}_1)^{-1}$ to bring the inverse $R$-matrix
in the same form as the original one:
\begin{equation}
  \label{eq:53}
   (\mathcal{R}_0^{\mathrm{op}})^{-1}
  = (\mathcal{R}_0^{\mathrm{op}})^{-1} \exp \left[ \sum_{n \geq 1}
    \frac{n}{\kappa_n}  
      \gamma^{-\frac{n}{2}} H^{-}_{-n}\otimes \gamma^{\frac{n}{2}}
      H^{+}_n\right].
\end{equation}
Since we are computing the vacuum matrix element, the $\mathcal{R}_2$
can be ignored, so after substituting $H^{\pm}_n$ from
Eqs.~\eqref{eq:31}--~\eqref{eq:56} we have
\begin{equation}
  \label{eq:51}
  \exp \left[ \sum_{n \geq 1}
    \frac{n}{\kappa_n}  
      \gamma^{-\frac{n}{2}} H^{-}_{-n}\otimes \gamma^{\frac{n}{2}}
      H^{+}_n\right]
 ( |\varnothing,w\rangle_{\mathcal{F}^{(0,1)}_{q,t^{-1}}(w)} \otimes
  \ldots) = \exp \left[ \sum_{n \geq 1} \frac{w^{-n}}{n} \frac{\left(1-
        (t/q)^n \right)}{(1-q^{-n})} a_n \right].
\end{equation}
Combining the pieces of the inverse $R$-matrix together we get
\begin{equation}
  \label{eq:57}
  \bar{\chi}_{q,t^{-1}}^{q,t^{-1}}(u,w) = \left(
    \frac{t}{q} \right)^{\frac{Q}{2}} \exp \left[ \sum_{n \geq 1}
    \frac{w^{-n}}{n} \frac{\left( 1- (t/q)^n \right)}{1-q^{-n}} a_n \right]
\end{equation}

We find that the formulas~\eqref{eq:18} indeed match the
answers~\eqref{eq:34} obtained from multiplication of a
pair of intertwining operators. We have thus verified that our
definition of the crossing operator is self-consistent: the degenerate
resolution~\eqref{eq:16} with two branes meeting in the same constant
$t$ plane indeed reproduces the universal $R$-matrix and hence
satisfies the Yang-Baxter equation. Of course, the Yang-Baxter
equation for such crossings can also be checked directly as was done
in~\cite{Awata:2016mxc}, \cite{Awata:2016bdm}.

\subsection{Functoriality of the crossing}
\label{sec:funct-cross}
Many interesting properties follow from Eqs.~\eqref{eq:18}. For example, suppose that we have \emph{a pair} of NS5
branes intersecting a single D5 brane:
\begin{equation}
  \label{eq:38}
      \includegraphics[valign=c]{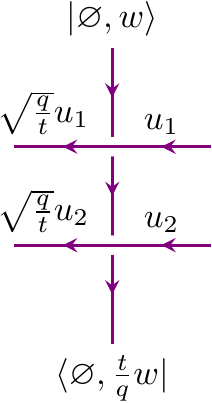} \quad =
      \quad (\mathcal{R}_{12} \mathcal{R}_{23})|_{\mathcal{F}^{(0,1)}_{q,t^{-1}}(w) \otimes
    \mathcal{F}^{(0,1)}_{q,t^{-1}}(u_1) \otimes
    \mathcal{F}^{(0,1)}_{q,t^{-1}}(u_2)}
\end{equation}
The corresponding combination of operators is a product of two
universal $R$-matrices contracted over one of their legs. We can
notice that this is precisely the combination appearing in one of the
axioms of the universal $R$-matrix:
\begin{equation}
  \label{eq:39}
  ( \Delta \otimes 1) \mathcal{R} = \mathcal{R}_{12} \mathcal{R}_{23}.
\end{equation}
Thus, the operator corresponding to the double crossing~\eqref{eq:38}
can be found by taking the coproduct of $\mathcal{R}_0$ and the
generators $H_{-n}^{-}$ inside the exponent in
$\mathcal{R}_1$~\eqref{eq:27}. Notice that the argument from
sec.~\ref{sec:lax-matrix-from-1} about the vanishing of $\mathbb{R}_2$
is still valid because it has to do with the vacuum states in the
\emph{vertical} representation on which the coproduct doesn't
act. The formulas for the DIM coproduct $\Delta$ can be found in
Appendix~A.2 of~\cite{Zenkevich:2018fzl}. In particular we have
\begin{align}
  \label{eq:2}
  \Delta(H_n^{\pm}) &= H^{\pm}_n\otimes \gamma^{\mp \frac{n}{2}} + \gamma^{\pm \frac{n}{2}} \otimes H^{\pm}_n,\\ 
  \Delta(\gamma) &= \gamma \otimes \gamma. \label{eq:4}
\end{align}
Applying Eqs.~\eqref{eq:2}, \eqref{eq:4} we get:
\begin{multline}
  \label{eq:40}
  (\mathcal{R}_{12}
  \mathcal{R}_{23})|_{\mathcal{F}^{(0,1)}_{q,t^{-1}}(w) \otimes
    \mathcal{F}^{(0,1)}_{q,t^{-1}}(u_1) \otimes
    \mathcal{F}^{(0,1)}_{q,t^{-1}}(u_2)}= ((\Delta \otimes
  1)\mathcal{R})|_{\mathcal{F}^{(0,1)}_{q,t^{-1}}(w) \otimes
    \mathcal{F}^{(0,1)}_{q,t^{-1}}(u_1) \otimes
    \mathcal{F}^{(0,1)}_{q,t^{-1}}(u_2)} =\\
  = \left( \frac{q}{t} \right)^{\frac{Q_1 + Q_2}{2}} \exp \left[
        -\sum_{n \geq 1} \frac{w^n}{n} \frac{(1-(t/q)^n)}{1-q^n}
        \left( a^{(1)}_{-n} + \left(\frac{t}{q}\right)^{\frac{n}{2}}
          a^{(2)}_{-n} \right) \right]=\\
      = \chi^{q,t^{-1}}_{q,t^{-1}}(u_1,w) \otimes
      \chi^{q,t^{-1}}_{q,t^{-1}}\left(u_2, \sqrt{\frac{t}{q}} w
      \right) 
\end{multline}
where $Q_{1,2}$ and $a^{(1,2)}_n$ are the shift operators and
Heisenberg generators acting on the first and second horizontal Fock
spaces respectively. The two factors in the last line of
Eq.~\eqref{eq:40} are completely factorized. We have thus proven that
the double crossing with vacuum states at both ends is just a product
of two vacuum crossings:
\begin{equation}
  \label{eq:41}
\boxed{\quad \includegraphics[valign=c]{figures/crossing-1-crop} \quad
  =  \quad   \includegraphics[valign=c]{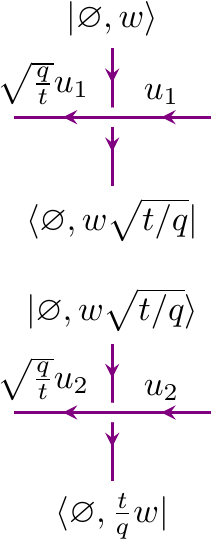} \quad}
\end{equation}
In other words, no nontrivial states are allowed on the segment of the
vertical line between between the crossings.  This phenomenon is in
fact well-known in the refined topological vertex formalism:
degenerate crossings force the vacuum state to ``pass through'' the
crossings downwards. This can also be verified directly using the more
direct approach of sec.~\ref{sec:lax-matrix-from}.

The factorization argument is general: for any number of horizontal
legs crossing a vertical line the resulting operator can be found by
taking the coproduct multiple times. The result will be of the form
\begin{equation}
  \label{eq:8}
\includegraphics[valign=c]{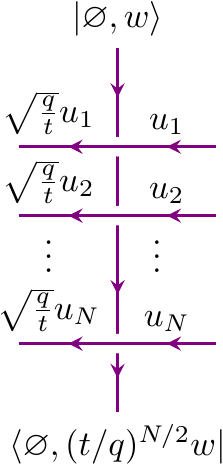} \quad  =  \quad  \chi^{q,t^{-1}}_{q,t^{-1}}(u_1,w) \otimes
  \chi^{q,t^{-1}}_{q,t^{-1}}\left(u_2, (t/q)^{\frac{1}{2}} w
  \right) \otimes \cdots \otimes \chi^{q,t^{-1}}_{q,t^{-1}}\left(u_N, (t/q)^{\frac{N-1}{2}} w
  \right)
\end{equation}
Again, this can also be seen using the refined topological vertices
$\Psi$ and $\Psi^{*}$ as in sec.~\ref{sec:lax-matrix-from}.

\subsection{Crossing between two 5-branes of different types}
\label{sec:cross-two-diff}
The next logical step in our exploration of the brane-algebra
dictionary is to consider a crossing of a D5 and NS5 branes sharing
only a single $\mathbb{C}$ factor instead of a $\mathbb{C}^2$
plane. Without loss of generality we can consider the crossing between
an NS5$_{q,t^{-1}}$ brane and a D5$_{q,t/q}$ brane:
\begin{equation}
  \label{eq:42}
  X^{q,t/q}_{q,t^{-1}}(u,w) \quad =
  \quad \includegraphics[valign=c]{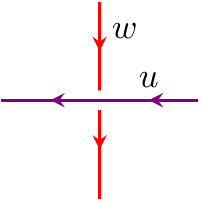}
  \qquad \qquad \qquad  \bar{X}^{q,t/q}_{q,t^{-1}}(u,w) \quad =
  \quad \includegraphics[valign=c]{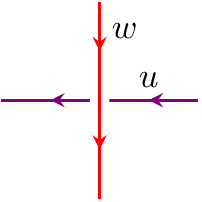}
\end{equation}

Here we don't have the opportunity to deform the picture so that the
two branes are brought into the same plane of constant $t$ and merged
into a web. However, the crossing should still satisfy the Yang-Baxter
equation hence should be a DIM Lax operator. It should act in the
tensor product of the vertical Fock space
$\mathcal{F}^{(0,1)}_{q,t/q}(w)$ corresponding to D5$_{q,t/q}$ and the
horizontal Fock space $\mathbb{F}^{(1,0)}_{q,t^{-1}}(u)$,
corresponding to NS5$_{q,t^{-1}}$. Therefore, we can write similarly
to~\eqref{eq:20} (notice the difference in the indices of the Fock
representations):
\begin{equation}
  \label{eq:43}
      \boxed{\begin{aligned}
  X^{q,t/q}_{q,t^{-1}}(u,w) & = \mathcal{R}|_{\mathcal{F}^{(0,1)}_{q,t/q}(w) \otimes
    \mathcal{F}^{(0,1)}_{q,t^{-1}}(u)},\\
  \bar{X}^{q,t/q}_{q,t^{-1}}(u,w) &= 
  \mathcal{R}^{-1}|_{\mathcal{F}^{(0,1)}_{q,t/q}(w)
    \otimes \mathcal{F}^{(0,1)}_{q,t^{-1}}(u)}.% \label{eq:19}
  \end{aligned}}
\end{equation}
Using the same arguments as in sec.~\ref{sec:inters-5-bran} we can
evaluate the vacuum matrix elements of the Lax operators~\eqref{eq:43}
explicitly:
\begin{multline}
  \label{eq:44}
  \chi_{q,t^{-1}}^{q,t/q}(u,w) = ( \ldots \otimes \langle \varnothing,
  \sqrt{t/q} \, w | ) X_{q,t^{-1}}^{q,t/q}(u,w)(|\varnothing,w\rangle
  \otimes \ldots) \quad
  =\quad \includegraphics[valign=c]{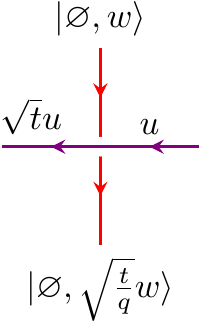} \quad
  = \\
  =t^{\frac{Q}{2}} \exp \left[ -\sum_{n \geq 1} \frac{w^n}{n}
    \frac{\left(1- t^{-n} \right)}{(1-q^n)} a_{-n} \right],
\end{multline}
    \begin{multline}
    \label{eq:45}
    \bar{\chi}_{q,t^{-1}}^{q,t/q}(u,w) =
  (  \ldots \otimes \langle \varnothing, \sqrt{q/t} \, w | )
  X_{q,t^{-1}}^{q,t/q}(u,w)(|\varnothing,w\rangle \otimes \ldots)\quad
  =\quad  \includegraphics[valign=c]{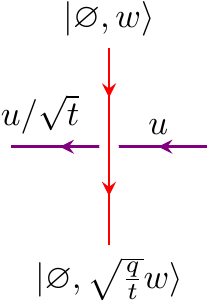}\quad=\\
  =t^{-\frac{Q}{2}} \exp \left[ -\sum_{n \geq 1} \frac{w^{-n}}{n} \left( \frac{t}{q} \right)^n
    \frac{\left(1- t^n \right)}{(1-q^{-n})} a_n \right]
\end{multline}

\subsection{Hanany-Witten D3 brane creation}
\label{sec:hanany-witten-brane}
There is something curious about the
operators~\eqref{eq:44}~\eqref{eq:45}, since they very much resemble
pieces of the D3$_q$-NS5$_{q,t^{-1}}$ junction~\eqref{eq:10}, which
reads~\cite{Zenkevich:2018fzl}
\begin{equation}
  \label{eq:46}
   \Phi^q_{q,t^{-1}} (|w\rangle \otimes \ldots)  =
   \quad  \includegraphics[valign=c]{figures/vect-vert-color-simpl-crop}
   \quad = t^Q w^{\frac{\ln u}{\ln q}} \exp \left[ - \sum_{n\geq 1} \frac{w^n}{n}
    \frac{1-t^{-n}}{1-q^n} a_{-n}  \right] \exp \left[ \sum_{n\geq 1} \frac{w^{-n}}{n}
    \frac{1-t^n}{1-q^{-n}} a_n \right]
\end{equation}
Comparing Eqs.~\eqref{eq:44}, \eqref{eq:45} with Eq.~\eqref{eq:46} we
arrive at a remarkable formula:
\begin{equation}
  \label{eq:47}
  \chi_{q,t^{-1}}^{q,t/q}(u,w) = w^{-\frac{\ln u}{\ln q} + \frac{\ln
    t}{2 \ln q}} \Phi^q_{q,t^{-1}} \left( |w\rangle \otimes
  \bar{\chi}_{q,t^{-1}}^{q,t/q}\left(u, \frac{t}{q} w \right)\right),
\end{equation}
which can be put in pictorial form (we omit the scalar prefactor in
the pictures)
\begin{equation}
  \label{eq:58}
    \includegraphics[valign=c]{figures/d5-ns5-hw-vac-crop} \quad
  = \quad  \includegraphics[valign=c]{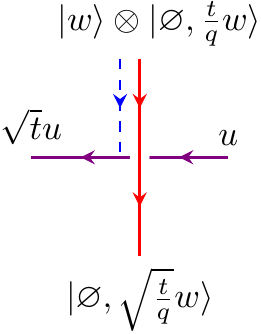}
\end{equation}
Furthermore, we can make use of the D3-D5 junction~\eqref{eq:13} to
write
\begin{equation}
  \label{eq:48}
  |w\rangle_{\mathcal{V}_q} \otimes \left|\varnothing,
  \frac{t}{q} w \right\rangle_{\mathcal{F}^{(0,1)}_{q,t/q}(wt/q)} =
  \tilde{\Phi}^{*q}_{q,t/q} |\varnothing, w\rangle.
\end{equation}
Eq.~\eqref{eq:48} and the identity~\eqref{eq:47} can be combined
compactly into the following picture:
\begin{equation}
  \label{eq:49}
 \boxed{\quad \includegraphics[valign=c]{figures/d5-ns5-hw-vac-crop} \quad
  = \quad   \includegraphics[valign=c]{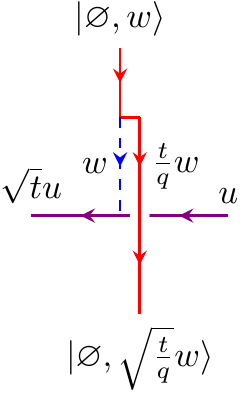}\quad }
\end{equation}
The picture~\eqref{eq:49} is the algebraic incarnation of the
Hanany-Witten brane creation effect~\cite{Hanany:1996ie}. When a
D5$_{q,t/q}$ brane is moved past an NS5$_{q,t^{-1}}$ a D3$_q$ brane is
created, which is stretched between the D5 and NS5 branes. This brane
move in our language is an equivalence transformation of the network
of intertwiners: the networks before and after the move coincide as
operators acting between DIM representations.

\subsection{Crossing of a NS5 brane and a D3 brane}
\label{sec:crossing-vector-fock}
Having understood the general pattern --- that brane crossings
correspond to DIM Lax operators in various representations --- we can
consider more examples of such operators. Perhaps the simplest example
is the D3 brane crossing an NS5$_{q,t^{-1}}$ brane. Here there are two
possibilities: either the D3 brane is D3$_q$ or D3$_{t^{-1}}$ and
shares a $\mathbb{C}$ worldvolume factor with the NS5$_{q,t^{-1}}$
brane, or it is the D3$_{t/q}$ brane and is completely orthogonal to
the NS$_{q,t^{-1}}$ brane. We consider these possibilities in turn.

\subsubsection{NS5$_{q,t^{-1}}$ and D3$_q$ branes.}
\label{sec:ns5_q-t-1}
For all the crossings we will be interested in the analogues of the
vacuum matrix elements of the corresponding Lax operators. There is no
vacuum state in the vector representation, so we are going to use the
ket state with fixed vertical position $|w\rangle$ instead of it. The
relevant bra state is $\langle \sqrt{\frac{t}{q}} w|$ because the
shift in the spectral parameter of the vertical brane is governed by
the central charge of the horizontal brane, which is
$\sqrt{\frac{t}{q}}$ in the present case.

We the crossing operators for NS5$_{q,t^{-1}}$ and D3$_q$ branes
$x^q_{q,t^{-1}}$ and $\bar{x}^q_{q,t^{-1}}$ since they are ``smaller
cousins'' of $\chi$ and $\bar{\chi}$ operators from
sec.~\ref{sec:degen-resolv-conif}. As in
sec.~\ref{sec:degen-resolv-conif},~\ref{sec:lax-matrix-from-1} there
are two ways to compute the crossings: using a pair of intertwiners
$\Phi$, $\Phi^{*}$, or from the universal $R$-matrix. The same results
agree. We write here

Plugging the expressions for $H_n^{\pm}$ in the vector representation $\mathcal{V}_q$
\begin{align}
  \label{eq:59}
  H_n^{+} |w\rangle_{\mathcal{V}_q} &=
    \frac{w^n(1-t^{-n}) \left( 1 - \left( \frac{t}{q} \right)^n \right)}{n}|w\rangle_{\mathcal{V}_q},\\
    H_n^{-} |w\rangle_{\mathcal{V}_q} &= \frac{w^{-n} (1-t^n) \left( 1
        - \left( \frac{q}{t} \right)^n \right)}{n}
    |w\rangle_{\mathcal{V}_q},\label{eq:5}
\end{align}
into the universal $R$-matrix~\eqref{eq:21} we get the crossing
operators:
\begin{align}
  \label{eq:60}
  x^q_{q,t^{-1}} (u,w)&=
  \quad  \includegraphics[valign=c]{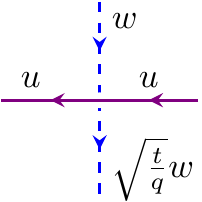}\quad =\exp \left[ - \sum_{n \geq 1} \frac{w^n}{n}
    \frac{(1-t^{-n}) \left( 1 - \left( \frac{t}{q} \right)^n
      \right)}{(1-q^n)} a_{-n}  \right], \\
  \bar{x}^q_{q,t^{-1}} (u,w)&=
  \quad  \includegraphics[valign=c]{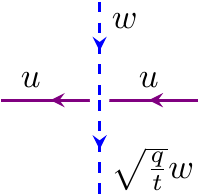}\quad = \exp \left[ \sum_{n \geq 1} \frac{w^{-n}}{n}
    \frac{(1-t^n) \left( 1 - \left( \frac{t}{q} \right)^n
      \right)}{(1-q^n)} a_n  \right].\label{eq:61}
\end{align}
Notice that the spectral parameter of the NS5 brane is unchanged at
the crossing with a D3 brane.

\subsubsection{NS5$_{q,t^{-1}}$ and D3$_{t/q}$ branes.}
\label{sec:ns5_q-t-1-1}
In this case the pair of branes shares only the $S^1$ direction. As in
the previous section we can use the expressions for $H^{\pm}_n$ in the
vector representation $\mathcal{V}_{t/q}$ (obtained from
Eqs.~\eqref{eq:59}, \eqref{eq:5} by the exchange $q \leftrightarrow
t/q$) to get the following crossing operators:
\begin{align}
  \label{eq:55}
  x^{t/q}_{q,t^{-1}} (u,w)&\quad =
  \quad  \includegraphics[valign=c]{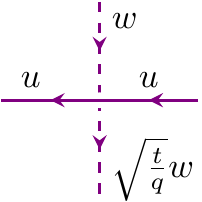}\quad = \quad\exp \left[ - \sum_{n \geq 1} \frac{w^n}{n}
    (1-t^{-n})
       a_{-n}  \right], \\
  \bar{x}^{t/q}_{q,t^{-1}} (u,w)&\quad =
  \quad  \includegraphics[valign=c]{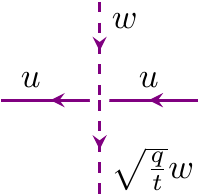}\quad
  = \quad \exp \left[ - \sum_{n \geq 1} \frac{w^{-n}}{n}
    \left( \frac{t}{q} \right)^n (1-t^n) a_n  \right].\label{eq:62}
\end{align}

\subsection{Hanany-Witten $(p,q)$-string creation}
\label{sec:hanany-witten-p}
We notice that the crossing operators~\eqref{eq:55}, \eqref{eq:62}
between orthogonal NS5 and D3 branes have particularly simple
form. Following the example of sec.~\ref{sec:hanany-witten-brane} we
compare $x^{t/q}_{q,t^{-1}}$ with $\bar{x}^{t/q}_{q,t^{-1}}$ and find
that they are related using the DIM generating currents $x^{\pm}(z)$
(see Appendix A of~\cite{Zenkevich:2018fzl} for the complete list of
definitions). In the Fock representation
$\mathcal{F}_{q,t^{-1}}^{(1,0)}(u)$  the generating currents of DIM act as follows:
\begin{align}
  \label{eq:63}
  x^{+}(z)|_{\mathcal{F}_{q,t^{-1}}^{(1,0)}(u)} &=
    \frac{u}{(1-q^{-1})(1-t)} \exp \left[ \sum_{n \geq 1}
      \frac{1-t^{-n}}{n} a_{-n}w^n \right] \exp \left[ - \sum_{n
        \geq 1} \frac{1-t^n}{n} a_nw^{-n} \right],\\
    x^{-}(z)|_{\mathcal{F}_{q,t^{-1}}^{(1,0)}(u)} &=
    \frac{u^{-1}}{(1-q)(1-t^{-1})} \exp \left[ -\sum_{n \geq 1}
      \frac{1-t^{-n}}{n} \left( \frac{t}{q} \right)^{\frac{n}{2}}
      a_{-n}w^n \right] \exp \left[ \sum_{n \geq 1} \frac{1-t^n}{n}
      \left( \frac{t}{q} \right)^{\frac{n}{2}} a_nw^{-n}
    \right], \label{eq:64}
  \end{align}
Using the formulas~\eqref{eq:63}--\eqref{eq:64} we can write the
following identity:
\begin{equation}
  \label{eq:76}
  \bar{x}^{t/q}_{q,t^{-1}} (u,w) = \frac{(1-q^{-1})(1-t)}{u} x^{t/q}_{q,t^{-1}} \left( u,
    \frac{q}{t} w \right) x^{+}\left( \frac{q}{t} w \right)
\end{equation}
Now we notice that the shift in the argument $w \to \frac{q}{t} w$ in
the r.h.s. of Eq.~\eqref{eq:76} can be reproduced by the action of
$x^{-}(z)$ in the \emph{vector} representation. More explicitly, the
action of generating currents in representation $\mathcal{V}_{t/q}$
reads
\begin{align}
  x^{+}(z)|w\rangle_{\mathcal{V}_{t/q}} &= - \frac{1}{1-q/t} \delta
    \left( \frac{w}{z} \right) \left| \frac{t}{q} w
    \right\rangle_{\mathcal{V}_{t/q}},\label{eq:66}\\
    x^{-}(z)|w\rangle_{\mathcal{V}_{t/q}} &= - \frac{1}{1-t/q} \delta
    \left( \frac{q}{t}\frac{w}{z} \right) \left| \frac{q}{t} w
    \right\rangle_{\mathcal{V}_{t/q}}.\label{eq:65}
\end{align}
Taking into account the delta-function in Eq.~\eqref{eq:65} we can
write Eq.~\eqref{eq:76} in a nice integral form:
\begin{equation}
  \label{eq:81}
  \bar{x}^{t/q}_{q,t^{-1}} (u,w) = -
  \frac{(1-q^{-1})(1-t)(1-t/q)}{u}\oint \frac{dz}{z} \left(
  \ldots \otimes \langle \sqrt{t/q} w | \right) \mathcal{R} \left(x^{-}(z) |w\rangle \otimes x^{+}\left( z \right)\right).
\end{equation}
The identity~\eqref{eq:81} seems complicated but the corresponding
picture of intertwiners is a lot more elucidating:
\begin{equation}
  \label{eq:82}\boxed{
 \quad \includegraphics[valign=c]{figures/d3-ns5-diff-inv-crop}\quad =
  \quad \sum_{r \in \mathbb{Z}} \quad  \includegraphics[valign=c]{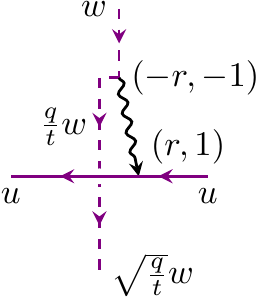}\quad}
\end{equation}
The wavy arrow denotes the $(p,q)$-string, as described in
sec.~\ref{sec:d1-joins-d3}. The beginning of the arrow puts DIM
generator $P_{(-r,-1)}$ on the corresponding brane, while the endpoint
of the arrow inserts $P_{(r,1)}$. The resulting sum over $r$ is the
same as the integral over $z$ in Eq.~\eqref{eq:81}.

Eq.~\eqref{eq:82} is nothing but another instance of the Hanany-Witten
effect: when a D3 brane passes through an orthogonal NS5 brane a
$(p,q)$-string is created. But it cannot be just a single
$(p,q)$-string for the following reason. The NS5 brane is invariant
under a subgroup of the $S$-duality group generated by an element $STS
\in SL(2,\mathbb{Z})$, and D3 brane is invariant under the whole
$SL(2,\mathbb{Z})$. Therefore the $(p,q)$-brane state should also be
invariant, and this is achieved by taking the linear superposition of
states with all possible $r$.

\subsection{Crossing of Fock and MacMahon representations}
\label{sec:cross-with-macm}
We can also write the crossing operators between an NS5 brane and
whatever brane corresponds to a vertical MacMahon representation
$\mathcal{M}^{(0,c)}(w)$. The action of DIM on
$\mathcal{M}^{(0,c)}(w)$ is given by~\cite{FJMM-Bethe-2}:
\begin{align}
  \label{eq:77}
    H^{+}_n|\varnothing, w \rangle_{\mathcal{M}^{(0,c)}(w)} &=
    \frac{w^n}{n} \left( 1 - K^n \right)|\varnothing,w
    \rangle_{\mathcal{M}^{(0,c)}(w)},\\
    H^{-}_{-n}|\varnothing,w
    \rangle_{\mathcal{M}^{(0,c)}(w)} &= \frac{w^{-n}}{n}
    \left( 1 - K^{-n} \right)|\varnothing,w
    \rangle_{\mathcal{M}^{(0,c)}(w)},\label{eq:78}
\end{align}
where $K=e^c$ parametrizes the central charge of the
representation. We have
\begin{align}
  \label{eq:79}
  \includegraphics[valign=c]{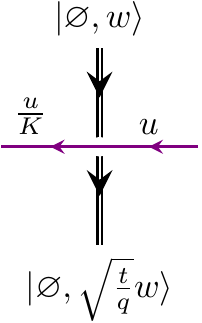}\quad &= \quad K^{-Q} \exp \left[ -\sum_{n \geq 1} \frac{w^n}{n}
    \frac{\left(1- K^n \right)}{(1-q^n)} a_{-n} \right],\\
    \includegraphics[valign=c]{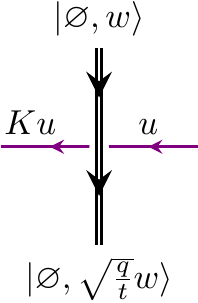}\quad &= \quad K^Q \exp \left[ -\sum_{n \geq 1} \frac{w^{-n}}{n}
    \left( \frac{t}{q} \right)^n \frac{\left(1- K^{-n} \right)}{(1-q^{-n})} a_n \right],\label{eq:80}
\end{align}
where the double line denotes the MacMahon representation. We have
discussed some hints about the role of MacMahon modules in the
Introduction, but this part in our brane/algebra dictionary is still
not clear.

\section{Higgsed vertices from refined topological vertices}
\label{sec:higgs-vert-from}

In this section we explain how obtain the higgsed intertwiners
of~\cite{Zenkevich:2018fzl} from the intertwiners of~\cite{AFS}. In
more physical terms we show how to obtain D3 brane junctions from the
junctions involving only 5-branes.

In sec.~\ref{sec:degen-resolv-conif} we have argued that the operator
corresponding to crossing NS5$_{q,t^{-1}}$ and D5$_{q,t^{-1}}$ branes
can be obtained from a degenerate resolved picture, where two
5-brane junctions are placed ``near each other'':
\begin{equation}
  \label{eq:83}
  C(u,w, (q/t)^{\frac{1}{2}}) \quad =
  \quad \includegraphics[valign=c]{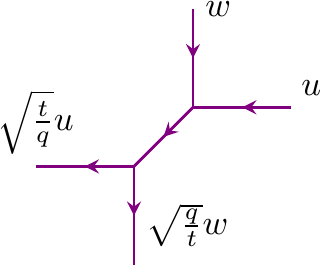}\quad =
  \quad \includegraphics[valign=c]{figures/ns5-d5-simpl-simpl-crop}
  \quad  =\quad  \bar{X}_{q,t^{-1}}^{q,t^{-1}}(u,w). 
\end{equation}
We would like to understand what happens if we tune the parameters of
the brane web in the l.h.s. to another value, with the parameter of
the resolution being $Q = \sqrt{\frac{t}{q}} t$. Making an explicit
computation with intertwining operators $\Psi$ and $\Psi^{*}$ one
finds that the resulting operator is a product of two: the crossing
$\bar{X}_{q,t^{-1}}^{q,t^{-1}}(u, t w)$ and the higgsed intertwiner
$\Phi^q_{q,t^{-1}}$:
\begin{equation}
  \label{eq:84}
  C(u,w, (q/t)^{\frac{1}{2}}t^{-1}) (|\lambda, w\rangle_{\mathcal{F}^{(0,1)}_{q,t^{-1}}(w)} \otimes \ldots)  =  
  \Phi^q_{q,t^{-1}}(|q^{\lambda_1}w\rangle_{\mathcal{V}_q} \otimes \bar{X}_{q,t^{-1}}^{q,t^{-1}}(u,
  t^{-1}w) (|\lambda\backslash \lambda_1,
  w\rangle_{\mathcal{F}^{(0,1)}_{q,t^{-1}}(w)} \otimes \ldots)). 
\end{equation}
We can now use the intertwiner from sec.~\ref{sec:d3-brane-joins} to
merge the states $|q^{\lambda_1}\rangle$ and $|\lambda\backslash
\lambda_1, w\rangle_{\mathcal{F}^{(0,1)}_{q,t^{-1}}(w)}$. We can then
draw the following picture of intertwiners:
\begin{equation}
  \label{eq:85}
\boxed{\quad  \includegraphics[valign=c]{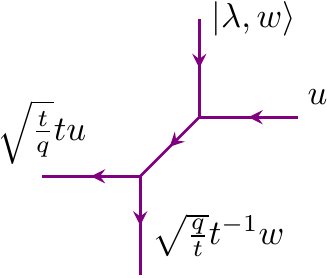}\quad =
  \quad \includegraphics[valign=c]{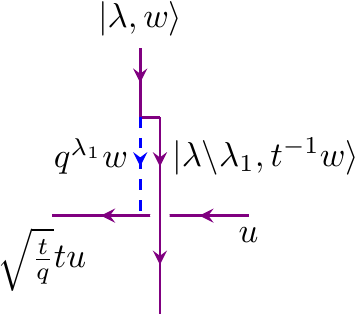}\quad}
\end{equation}
The identity~\eqref{eq:85} is the reason why the higgsing procedure
works: one can tune the parameters of the 5-brane web so that a D3
brane appears. Notice that Eq.~\eqref{eq:85} identity differs in an
essential way from Hanany-Witten identity Eq.~\eqref{eq:49}, where the
5-branes, being orthogonal to each other, cannot
merge. Since~\eqref{eq:85} is an operator identity, one can use it as
a part of any networks of intertwiners --- the identity applies
locally.

One more useful identity when transforming 5-brane webs into higgsed
networks is the following brane move, which follows directly from the
functoriality of crossing operators:
\begin{equation}
  \label{eq:86}
  \boxed{\quad  \includegraphics[valign=c]{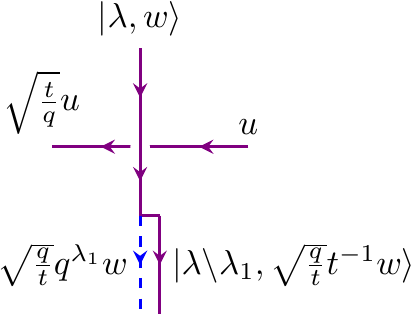}\quad =
  \quad \includegraphics[valign=c]{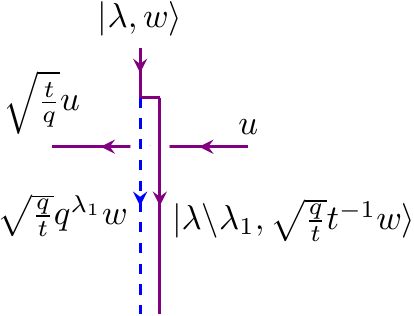}\quad }
\end{equation}
In the next section will see how using the identities~\eqref{eq:85}
and~\eqref{eq:86} one can transform a web of 5-branes with tuned
parameters into a higgsed network of intertwiners.

\section{$T[SU(N)]$ vs $FT[SU(N)]$ partition functions and brane
  factors}
\label{sec:tsun-ftsun-partition}
In this section we will study a nontrivial example of a $3d$ quiver
gauge theory, which can be obtained from our brane constructions. The
theory is denoted by $T[SU(N)]$ it is a linear quiver with gauge
groups $U(1) \times U(2) \times \ldots \times U(N-1)$ and a frozen
$U(N)$ flavour symmetry at one end. This theory can be thought of as
counting maps from a $3d$ space into full flag variety
$\mathrm{Fl}(1,2,3,\ldots, N)$. The theory has $\mathcal{N}=4$
supersymmetry in flat space, but for nontrivial equivariant parameter
$t$ it is softly broken to $\mathcal{N}=2^{*}$. See
e.g.~\cite{Zenkevich:2017ylb} for the review of the properties of
$T[SU(N)]$ theory.

The picture of intertwiners corresponding to this theory is (we draw
all the examples for $N=2$ only, but our approach is completely
general):
\begin{equation}
  \label{eq:89}
  Z_{T[SU(2)]}(\vec{u}, \vec{w}, q, t) \quad = \quad \includegraphics[valign=c]{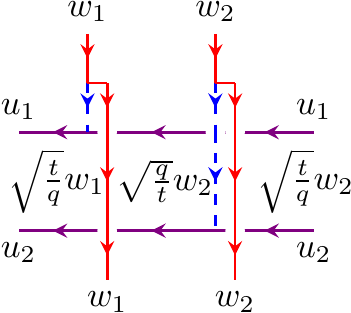}
\end{equation}

There is a close cousin of the $T[SU(N)]$ theory called\footnote{$F$
  stands for flip of a $\mathbb{C}$ direction spanned by the D5 branes
  --- from $\mathbb{C}^2_{q,t^{-1}}$ to $\mathbb{C}^2_{q,t/q}$.}
$FT[SU(N)]$ for which the red 5-branes in~\eqref{eq:89} are replaced
with violet ones:
\begin{equation}
  \label{eq:90}
  Z_{FT[SU(2)]}(\vec{u}, \vec{w}, q, t) \quad = \quad\includegraphics[valign=c]{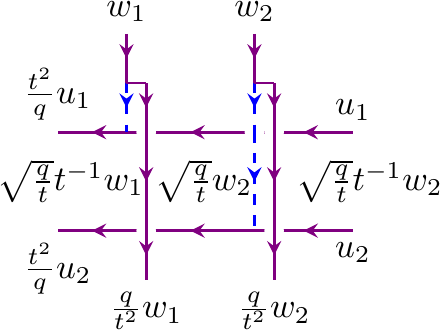}
\end{equation}
This theory even without equivariant deformation has only
$\mathcal{N}=2$ supersymmetry. However, it has almost the same
partition function as the $T[SU(N)]$ theory. Indeed the arrangement of
the D3 branes in both theories are the same, and so is the gauge
contribution to the partition function. The theories differ only by
what is usually called ``brane factors''~\cite{Zenkevich:2017ylb} ---
the contributions of the fields living in the worldvolume of the
5-branes. These fields are non-dynamical from the point of view of the
$3d$ theory and hence give a simple prefactor in front of the
partition function. We will explain the origin of the brane factors
algebraically in sec.~\ref{sec:comm-intertw}.

To do this we employ brane moves on both pictures~\eqref{eq:89},
\eqref{eq:90} first to send the D3-D5 junctions upwards to infinity,
so that the D3 and D5 branes can move independently. After that we
move D5 branes to the right of the picture, commuting them with all D3
branes that happen on the way. The commutation gives rise to the brane
factors. As soon as the D5 branes reach the vacuum state living on the
NS5 branes to the right of the picture they disappear. Indeed,
$\bar{\chi}$ from Eq.~\eqref{eq:34} as well as from Eq.~\eqref{eq:45}
contain only the annihilation operators $a_n$ with $n>0$. In this way
we are left with the partition function involving only D3 and NS5
branes, which we have already considered in~\cite{Zenkevich:2018fzl}.
  
\subsection{Crossings commute with screening charges}
\label{sec:cross-comm-with}
Consider two NS5$_{q,t^{-1}}$ branes with a segment of D3$_q$
stretched between them. Let there be also a D5$_{q,t^{-1}}$ brane
crossing underneath the NS5 branes:
\begin{equation}
  \label{eq:87}
  \includegraphics[valign=c]{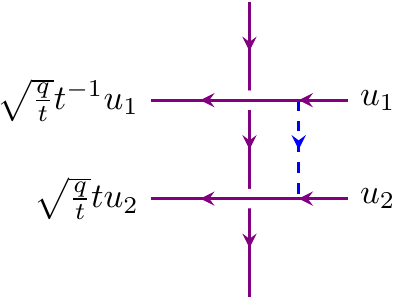}
\end{equation}

As pointed out in sec.~\ref{sec:funct-cross}, the crossings between
the D5 and the NS5 branes are described by a single universal
$R$-matrix. This $R$-matrix is an element of the tensor square of the
DIM algebra. The second factor of this tensor product acts on the
tensor product of two horizontal Fock representations. From this
seemingly abstract reasoning one can deduce that the D5 brane commutes
with the D3 brane. Indeed, as discussed in detail
in~\cite{Zenkevich:2018fzl} the finite D3 brane segment gives the
screening charge --- the intertwiner of a pair of Fock
representations. The screening charge commutes with any element of the
DIM algebra, hence with D5 crossing:
\begin{equation}
  \label{eq:88}
  \includegraphics[valign=c]{figures/move-3-crop}\quad = \quad \includegraphics[valign=c]{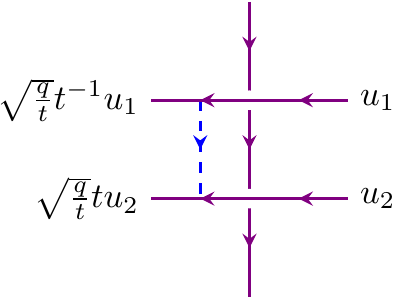}
\end{equation}
Of course this identity can also be checked by direct computation.

Due to the identity~\eqref{eq:88} when we commute the D5 brane
crossings with NS5-D3 junctions, we don't have to worry about the
\emph{intermediate} finite D3 brane segments. Only semi-infinite D3
branes will produce the brane factors.

\subsection{Brane factors from commutation relations between crossings
  and intertwiners}
\label{sec:comm-intertw}
In the previous section we have shown that the screening charges
corresponding to finite D3 brane segments commute with D5 branes
running beneath or above them. The situation is different for
semi-infinite D3 branes --- the commutation relations are nontrivial
for them. These commutation relations produce the brane factors for
the $3d$ theory. We thus have to take into account only the uppermost
NS5 brane on which the seim-infinite parts of D3 branes are
located. It is convenient at this point to write the D3-NS5 brane
crossings using D3-NS5 brane junctions.

Using the explicit expressions~\eqref{eq:18}, \eqref{eq:44},
\eqref{eq:45} we calculate the commutators:
\begin{multline}
  \bar{\chi}^{q,t^{-1}}_{q,t^{-1}}(z,u) \Phi^q_{q,t^{-1}}|w
  \rangle \otimes \ldots =  w^{\frac{1-\beta}{2}} \exp \left[ \sum_{n \geq 1}
    \frac{1}{n} \left( \frac{w}{z} \right)^n \frac{1-
      (q/t)^n}{1-q^n} \right] \Phi^q_{q,t^{-1}}|w
  \rangle \otimes \bar{\chi}^{q,t^{-1}}_{q,t^{-1}}(z,u) = \\
  =
  w^{\frac{1-\beta}{2}} \frac{\left(\frac{q}{t}\frac{w}{z} ;q \right)_{\infty} }{\left(\frac{w}{z} ;q \right)_{\infty} } \Phi^q_{q,t^{-1}}|w
  \rangle \otimes \bar{\chi}^{q,t^{-1}}_{q,t^{-1}}(z,u).\label{eq:91}
\end{multline}
\begin{multline}
  \bar{\chi}^{q,t/q}_{q,t^{-1}}(z,u) \Phi^q_{q,t^{-1}}|w
  \rangle \otimes \ldots =  w^{\frac{\beta}{2}} \exp \left[ \sum_{n \geq 1}
    \frac{1}{n} \left( \frac{w}{z} \right)^n \frac{1-
      t^n}{1-q^n} \right] \Phi^q_{q,t^{-1}}|w
  \rangle \otimes \bar{\chi}^{q,t^{-1}}_{q,t^{-1}}(z,u) = \\
  =
  w^{\frac{\beta}{2}} \frac{\left(\frac{w}{z} ;q
    \right)_{\infty} }{\left(t \frac{w}{z} ;q \right)_{\infty} } \Phi^q_{q,t^{-1}}|w
  \rangle \otimes \bar{\chi}^{q,t/q}_{q,t^{-1}}(z,u).\label{eq:92}
\end{multline}

We can then compute the matrix element of several ``dressed''
intertwiners $\Phi^q_{q,t^{-1}}(|w_i\rangle \otimes
\chi^{q,t^{-1}}_{q,t^{-1}}(t^{-1}w_i))$ (and
$\Phi^q_{q,t^{-1}}(|w_i\rangle \otimes
\chi^{q,t/q}_{q,t^{-1}}(w_it/q))$) in terms of the ``undressed'' ones
$\Phi^q_{q,t^{-1}}|w_i\rangle $:
\begin{multline}
  \langle \varnothing,  u_1 |
  \prod_{i=1}^n \Phi^q_{q,t^{-1}}|w_i\rangle \otimes \chi^{q,t^{-1}}_{q,t^{-1}}(t^{-1}w_i) |
  \varnothing,u_1 \rangle =\\
  = \prod_{i<j}^n \left[w_j^{\frac{1-\beta}{2}}
    \frac{\left(q\frac{w_j}{w_i} ;q \right)_{\infty} }{\left(t
        \frac{w_j}{w_i} ;q \right)_{\infty} } \right] \langle
  \varnothing, t^{\frac{3n}{2}}q^{-\frac{n}{2}} u_1 | \prod_{i=1}^n
  \Phi^q_{q,t^{-1}}|w_i\rangle \otimes | \varnothing, t^{\frac{n}{2}}q^{-\frac{n}{2}} u_1
  \rangle.\label{eq:96}
\end{multline}
\begin{multline}
  \langle \varnothing, t^{\frac{n}{2}} u_1 |
  \prod_{i=1}^n \Phi^q_{q,t^{-1}}|w_i\rangle \otimes
  \chi^{q,t/q}_{q,t^{-1}}\left(\frac{t}{q} w_i\right) |
  \varnothing,u_1 \rangle =\\
  = \prod_{i<j}^n \left[w_j^{\frac{\beta}{2}}
    \frac{\left(q \frac{w_j}{w_i} ;q \right)_{\infty} }{\left(\frac{q}{t}\frac{w_j}{w_i} ;q \right)_{\infty} } \right] \langle
  \varnothing, t^{\frac{n}{2}} u_1 | \prod_{i=1}^n
  \Phi^q_{q,t^{-1}}|w_i\rangle \otimes | \varnothing, t^{-\frac{n}{2}}u_1
  \rangle.\label{eq:97}
\end{multline}
Let us introduce the ``undressed'' partition function
\begin{equation}
  \label{eq:93}
  Z^{\mathrm{undressed}}_{T[SU(2)]}(\vec{u},\vec{w},q,t) \quad =
  \quad \includegraphics[valign=c]{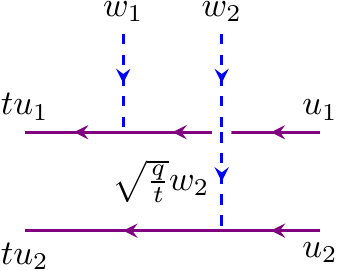}.
\end{equation}
By construction this partition function doesn't feel the difference
between the $T[SU(N)]$ and $FT[SU(N)]$ theories.

From~\eqref{eq:96} \eqref{eq:97} we can express two different dressed
partition functions through the undressed one and therefore get the
relation between the two:
\begin{align}
  \label{eq:95}
  Z_{T[SU(N)]}(\vec{u}, \vec{w}, q, t) &=  \prod_{i<j}^N \left[w_j^{\frac{\beta}{2}}
    \frac{\left(q \frac{w_j}{w_i} ;q \right)_{\infty}
    }{\left(\frac{q}{t}\frac{w_j}{w_i} ;q \right)_{\infty} } \right]
  Z^{\mathrm{undressed}}_{T[SU(2)]}(\vec{u},\vec{w},q,t),\\
  Z_{FT[SU(N)]}(\vec{u}, \vec{w}, q, t) &=  \prod_{i<j}^N \left[w_j^{\frac{1-\beta}{2}}
    \frac{\left(q \frac{w_j}{w_i} ;q \right)_{\infty}
    }{\left(t\frac{w_j}{w_i} ;q \right)_{\infty} } \right]
  Z^{\mathrm{undressed}}_{T[SU(2)]}(\vec{u},\vec{w},q,t).\label{eq:94}
\end{align}
Comparing Eqs.~\eqref{eq:95} and~\eqref{eq:94} we finally get the
relation between the $T[SU(N)]$ and $FT[SU(N)]$ partition functions:
\begin{equation}
  \label{eq:98}
\boxed{  Z_{FT[SU(N)]}(\vec{u}, \vec{w}, q, t) =  \prod_{i<j}^N
  \left[w_j^{\frac{1}{2} - 2\beta}
    \frac{\left( \frac{q}{t} \frac{w_j}{w_i} ;q \right)_{\infty}
    }{\left(t\frac{w_j}{w_i} ;q \right)_{\infty} } \right]Z_{T[SU(N)]}(\vec{u}, \vec{w}, q, t)}
\end{equation}
This coincides with the expressions for the flipping fields from
e.g.~\cite{Zenkevich:2017ylb}.

\subsection{$S$-duality and spectral duality}
\label{sec:s-duality-spectral}
$S$-duality of Type IIB theory in our formalism is just the rotation
of the picture by\footnote{Mirror image along the diagonal is
  essentially equivalent to the $\frac{\pi}{2}$ rotation in this
  setup.} $\frac{\pi}{2}$. In the algebraic language $S$-duality acts
on DIM algebra by automorphisms\footnote{DIM algebra is invariant
  under the automorphisms, but the coalgebra structure after an
  $SL(2,\mathbb{Z})$ transformations is twisted by a nontrivial
  Drinfeld cocycle. Thus $SL(2,\mathbb{Z})$ is an automorphism of DIM
  algebra as a Hopf algebra.}. Under this rotation the spectral
parameters $\vec{u}$ and $\vec{w}$ are exchanged. Using the brane
move~\eqref{eq:86} one can bring the $T[SU(N)]$ brane
picture~\eqref{eq:89} into the form symmetric under the mirror image
(also under $\frac{\pi}{2}$ rotation, but one needs to commute some
branes). However, there is an important caveat that the invariance
holds provided one exchanges the red and violet branes after the
rotation. This can be achieved by making the substitution $t
\leftrightarrow \frac{q}{t}$: DIM algebra is invariant and the only
things that change are the labels of its representations. We therefore
have for the $T[SU(N)]$ partition function:
\begin{equation}
  \label{eq:99}
  \boxed{Z_{T[SU(N)]}(\vec{u}, \vec{w}, q, t) = Z_{T[SU(N)]}\left(\vec{w},
    \vec{u}, q, \frac{q}{t} \right)}
\end{equation}

In $FT[SU(N)]$ theory the situation is different, since one doesn't
need to exchange any branes. So, $S$-duality of Type IIB implies:
\begin{equation}
  \label{eq:100}
    \boxed{Z_{FT[SU(N)]}(\vec{u}, \vec{w}, q, t) = Z_{FT[SU(N)]}(\vec{w},
    \vec{u}, q, t )}
\end{equation}
This duality can be called spectral, since it is related to the
spectral duality in integrable systems. The relation~\eqref{eq:98}
together with~\eqref{eq:99} and~\eqref{eq:100} implies a simple
transformation rule for $T[SU(N)]$ under the inversion $t
\leftrightarrow \frac{q}{t}$.

\subsection{Dualities in the double elliptic system}
\label{sec:dual-double-ellipt}
Let us also mention that the diagrams~\eqref{eq:89}, \eqref{eq:90} for
$T[SU(N)]$ and $FT[SU(N)]$ theories respectively can be compactified
in both directions. Naively, this gives rise to a $4d$ gauge theory
with affine quiver. In fact after the compactification the theory on
the D3 does not decouple from the sector living on the 5-branes, so
the situation is more involved. However, since our brane moves are
local we can still apply them to transform the brane diagram as we
wish. In this way it is possible to get from our picture a purely
5-brane diagram studied in~\cite{Fukuda:2020czf} in the context of
double elliptic (Dell) integrable systems.

It is remarkable that the duality properties of the compactified
partition function can be deduced from the same reasoning as before
the compactification. In particular one can compute the brane factors
relating the compactified $T[SU(N)]$ and $FTSU[N]$ theories. They are
expressed through double $q$-Pochhammer symbols.

\section{Conclusions and discussions}
\label{sec:conclusions}
In this paper we have enriched the dictionary between Type IIB brane
constructions and representations of DIM algebra. We have
systematically investigated different types of brane junctions and
crossings and found the corresponding intertwining operators. We have
argued that the brane crossing operators have a unified algebraic
meaning: they are given by the universal $R$-matrix for DIM algebra
taken in the representations associated with the pair of crossed
branes. We have found remarkable identities between crossing and
intertwining operators, which translate the Hanany-Witten brane
creation effect into the algebraic language. We have elucidated the
higgsing picture, i.e.\ the identity between 5-brane webs at certain
special points of their moduli spaces and brane networks with extra D3
branes. Finally, we have clarified the origin of $S$-duality and
spectral duality in $3d$ gauge theories living on D3 branes stretched
between 5-branes.

The structure of the DIM algebra and its representation ring turns out
to be so restrictive that it is tempting to propose that Type IIB
brane moves can be \emph{derived} from the DIM algebra. This would
mean that the theory (at least after the equivariant deformation) is
completely determined by its large hidden symmetry, just as the
properties of a $2d$ CFT can be derived just from knowing the Virasoro
algebra.

Though we had some success with the identification of algebraic
objects with branes, some points clearly deserve further
investigation. In particular, it is still not completely clear what is
the role of MacMahon representations. Another crucial point is the
coproduct in the DIM algebra whose role in the brane picture remains
mysterious. How does the DIM coproduct, which is not invariant under
$SL(2,\mathbb{Z})$ transformations arise from the branes? What is the
role of the $R$-matrix in the brane picture?  We leave all these
questions for the future.

In general, it would be also interesting to investigate the connection
between the universal $R$-matrices and the $R$-matrices obtained from
the tetrahedron equations as in~\cite{Bazhanov:2005as}. Their forms
are quite different: the former is the product over positive roots of
a root system, while the latter is a trace over an auxiliary
space. Nevertheless, they carry the same amount of information at
least in the $A_N$ case. It would also be interesting to obtain a
solution of the tetrahedron equation from which the DIM algebra
$R$-matrix can be obtained possibly along the lines
of~\cite{Gavrylenko:2020eov}.

As shown in~\cite{Zenkevich:2018fzl}, our approach reproduces the
matrix models computing holomorphic blocks of $3d$ supersymmetric
gauge theories. The brane pictures and rules we have considered here
and in~\cite{Zenkevich:2018fzl} can be understood from the purely
matrix model point of view. To this end one needs to ``project'' the
pictures on the $\mathbb{R}_t$. The D3 brane worldvolumes then become
segments separated by points, which represent 5-branes. The
integration over the positions of the intermediate D3 branes produces
the matrix model. This approach has been pursued e.g.\
in~\cite{Assel:2014awa}, where the action of $SL(2,\mathbb{Z})$ and
some brane rules for the matrix model were introduced. Our
$SL(2,\mathbb{Z})$ action on the DIM representation and intertwiners
matches that of~\cite{Assel:2014awa}, however we skip the details not
to further overcomplicate our presentation here.

Perhaps one could better understand the brane model we have introduced
using the $4d$ Chern-Simons theory~\cite{Costello} with DIM algebra
playing the role of the gauge group.

\section{Acknowledgements}
\label{sec:acknowledgements}
The author thanks M.~Bershtein, M.~Aganagic and C.~Schmid for
discussions. The author is supported by the RSF grant 18-71-10073.


\begin{thebibliography}{99}
\bibitem{Nakajima}
  H.~Nakajima, % Instantons on ALE spaces, quiver varieties, and Kac-Moody algebras,
  Duke Math.~J. 76(1994), no. 2, 365--416.\\
  H.~Nakajima, % Heisenberg algebra and Hilbert schemes of points on projective surfaces,
  Ann.~of Math. (2) 145 (1997), no. 2, 379--388.\\
  H.~Nakajima, % Quiver varieties and Kac-Moody algebras,
  Duke Math. J.91(1998), no. 3, 515--560.

\bibitem{Alday:2009aq} L.~F.~Alday, D.~Gaiotto and Y.~Tachikawa,
  %``Liouville Correlation Functions from Four-dimensional Gauge Theories,''
  Lett.\ Math.\ Phys.\ {\bf 91} (2010) 167
  doi:10.1007/s11005-010-0369-5 [arXiv:0906.3219 [hep-th]].\\
  N.~Wyllard,
  % \emph{$A_{N-1}$ conformal Toda field theory correlation functions
  % from conformal $N=2$ $SU(N)$ quiver gauge theories},
  JHEP {\bf 0911} (2009) 002, arXiv:0907.2189\\
  A.~Mironov and A.~Morozov, Nucl.\ Phys.\ {\bf B825} (2009) 1--37,
  arXiv:0908.2569

  
\bibitem{Zenkevich:2018fzl} Y.~Zenkevich,
% ``Higgsed network calculus,''
[arXiv:1812.11961 [hep-th]].

\bibitem{DIM} J. Ding, K.
  Iohara, %Generalization of Drinfeld quantum affine algebras,
  Lett. Math. Phys. {\bf 41} (1997) 181--193, q-alg/9608002\\
  K. Miki, J. Math. Phys. {\bf 48} (2007) 123520

    \bibitem{Awata:2005fa}
H.~Awata and H.~Kanno,
%``Instanton counting, Macdonald functions and the moduli space of D-branes,''
JHEP \textbf{05} (2005), 039
doi:10.1088/1126-6708/2005/05/039
[arXiv:hep-th/0502061 [hep-th]].

    \bibitem{Iqbal:2007ii}
  A.~Iqbal, C.~Kozcaz and C.~Vafa,
  %``The Refined topological vertex,''
  JHEP {\bf 0910} (2009) 069 doi:10.1088/1126-6708/2009/10/069
  [hep-th/0701156].


  \bibitem{AFS}
  H.~Awata, B.~Feigin and J.~Shiraishi,
  %``Quantum Algebraic Approach to Refined Topological Vertex,''
  JHEP {\bf 1203} (2012) 041
  doi:10.1007/JHEP03(2012)041
  [arXiv:1112.6074 [hep-th]].

\bibitem{Zenkevich:2019ayk}
Y.~Zenkevich,
%``$\mathfrak{gl}_N$ Higgsed networks,''
[arXiv:1912.13372 [hep-th]].

\bibitem{Kimura:2015rgi}
  T.~Kimura and V.~Pestun,
  %``Quiver W-algebras,''
  arXiv:1512.08533 [hep-th].\\
  T.~Kimura and V.~Pestun,
  %``Quiver elliptic W-algebras,''
  arXiv:1608.04651 [hep-th].\\
  T.~Kimura and V.~Pestun,
  %``Fractional quiver W-algebras,''
  arXiv:1705.04410 [hep-th].

  \bibitem{Gaiotto:2017euk}
  D.~Gaiotto and M.~Rapčák,
  %``Vertex Algebras at the Corner,''
  arXiv:1703.00982 [hep-th].\\
  T.~Procházka and M.~Rapčák,
  %``Webs of W-algebras,''
  JHEP {\bf 1811} (2018) 109
  doi:10.1007/JHEP11(2018)109
  [arXiv:1711.06888 [hep-th]].\\
  T.~Procházka and M.~Rapčák,
  %``$\mathcal{W}$-algebra Modules, Free Fields, and Gukov-Witten Defects,''
  arXiv:1808.08837 [hep-th].\\
  M.~Rapcak, Y.~Soibelman, Y.~Yang and G.~Zhao,
  %``Cohomological Hall algebras, vertex algebras and instantons,''
  arXiv:1810.10402 [math.QA].\\
  M.~Rap\v{c}\'ak,
%``On extensions of $ \mathfrak{gl}\widehat{\left(\left.m\right|n\right)} $ Kac-Moody algebras and Calabi-Yau singularities,''
JHEP \textbf{01} (2020), 042
doi:10.1007/JHEP01(2020)042
[arXiv:1910.00031 [hep-th]].\\
M.~Rapcak, Y.~Soibelman, Y.~Yang and G.~Zhao,
%``Cohomological Hall algebras and perverse coherent sheaves on toric Calabi-Yau 3-folds,''
[arXiv:2007.13365 [math.QA]].\\
D.~Gaiotto and M.~Rapcak,
%``Miura operators, degenerate fields and the M2-M5 intersection,''
[arXiv:2012.04118 [hep-th]].
  
\bibitem{Mironov:2016yue}
  A.~Mironov, A.~Morozov and Y.~Zenkevich,
  %``Ding–Iohara–Miki symmetry of network matrix models,''
  Phys.\ Lett.\ B {\bf 762} (2016) 196
  doi:10.1016/j.physletb.2016.09.033
  [arXiv:1603.05467 [hep-th]].\\
  H.~Awata, H.~Kanno, T.~Matsumoto, A.~Mironov, A.~Morozov, A.~Morozov, Y.~Ohkubo and Y.~Zenkevich,
  %``Explicit examples of DIM constraints for network matrix models,''
  JHEP {\bf 1607} (2016) 103
  doi:10.1007/JHEP07(2016)103
  [arXiv:1604.08366 [hep-th]].


\bibitem{Bourgine:2016vsq}
  J.~E.~Bourgine, M.~Fukuda, Y.~Matsuo, H.~Zhang and R.~D.~Zhu,
  %``Coherent states in quantum $\mathcal{W}_{1+\infty}$ algebra and qq-character for 5d Super Yang-Mills,''
  PTEP {\bf 2016} (2016) no.12,  123B05
  doi:10.1093/ptep/ptw165
  [arXiv:1606.08020 [hep-th]].\\
  J.~E.~Bourgine, M.~Fukuda, K.~Harada, Y.~Matsuo and R.~D.~Zhu,
  %``(p, q)-webs of DIM representations, 5d $ \mathcal{N}=1 $ instanton partition functions and qq-characters,''
  JHEP {\bf 1711} (2017) 034
  doi:10.1007/JHEP11(2017)034
  [arXiv:1703.10759 [hep-th]].\\
  J.~E.~Bourgine,
  %``Webs of Quantum Algebra Representations in 5d \({\mathcal {N}}=1\) Super Yang–Mills,''
  Springer Proc.\ Math.\ Stat.\  {\bf 263} (2017) 209.
  doi:10.1007/978-981-13-2715-5\_11\\
J.~E.~Bourgine, M.~Fukuda, Y.~Matsuo and R.~D.~Zhu,
  %``Reflection states in Ding-Iohara-Miki algebra and brane-web for D-type quiver,''
  JHEP {\bf 1712} (2017) 015
  doi:10.1007/JHEP12(2017)015
  [arXiv:1709.01954 [hep-th]].\\
  J.~E.~Bourgine and K.~Zhang,
  %``A note on the algebraic engineering of 4D $\mathcal{N}=2$ super Yang-Mills theories,''
  Phys.\ Lett.\ B {\bf 789} (2019) 610
  doi:10.1016/j.physletb.2018.11.066
  [arXiv:1809.08861 [hep-th]].\\
  J.~E.~Bourgine,
  %``Fiber-base duality from the algebraic perspective,''
  arXiv:1810.00301 [hep-th].

\bibitem{semi-inf} B.~Feigin, E.~Feigin, M.~Jimbo, T.~Miwa, E.~Mukhin, 
 %``Quantum continuous $\mathfrak{gl}_{\infty}$: Semi-infinite construction of representations,''
Kyoto J. Math. 51, no. 2 (2011), 337--364, [arXiv:1002.3100 [math.QA]]
  
\bibitem{Awata:2016mxc}
H.~Awata, H.~Kanno, A.~Mironov, A.~Morozov, A.~Morozov, Y.~Ohkubo and Y.~Zenkevich,
%``Toric Calabi-Yau threefolds as quantum integrable systems. $ \mathrm{\mathcal{R}} $ -matrix and $ \mathrm{\mathcal{R}}\mathcal{T}\mathcal{T} $ relations,''
JHEP \textbf{10} (2016), 047
doi:10.1007/JHEP10(2016)047
[arXiv:1608.05351 [hep-th]].

\bibitem{Awata:2016bdm}
H.~Awata, H.~Kanno, A.~Mironov, A.~Morozov, A.~Morozov, Y.~Ohkubo and Y.~Zenkevich,
%``Anomaly in RTT relation for DIM algebra and network matrix models,''
Nucl. Phys. B \textbf{918} (2017), 358-385
doi:10.1016/j.nuclphysb.2017.03.003
[arXiv:1611.07304 [hep-th]].

\bibitem{Awata:2018svb}
H.~Awata, H.~Kanno, A.~Mironov, A.~Morozov, K.~Suetake and Y.~Zenkevich,
%``The MacMahon $R$-matrix,''
JHEP \textbf{04} (2019), 097
doi:10.1007/JHEP04(2019)097
[arXiv:1810.07676 [hep-th]].

\bibitem{Negut-R} A.~Negut, % The R-matrix of the quantum toroidal
  % algebra,
[arXiv:2005.14182 [math.QA]]

\bibitem{FJMM-Bethe} B.~Feigin, M.~Jimbo, T.~Miwa, E.~Mukhin, % Finite
  % Type Modules and Bethe Ansatz for Quantum Toroidal
  % \({\mathfrak{gl}_1}\) .
  Commun. Math. Phys. 356, 285–327
  (2017). https://doi.org/10.1007/s00220-017-2984-9, [arXiv:1603.02765 [math.QA]]

  \bibitem{FJMM-Bethe-2} B.~Feigin, M.~Jimbo, T.~Miwa, E.~Mukhin, % Quantum toroidal gl(1) and Bethe ansatz,
  [arXiv:1502.07194 [math.QA]]

\bibitem{KhorTolst} S.~M.~Khoroshkin,
  V.~N.~Tolstoy, % Universal R-matrix for quantized (super)algebras
  Comm.~Math.~Phys.,  Volume 141, Number 3 (1991), 599--617.\\
      V.~N.~Tolstoy, S.~M.~Khoroshkin 
  % The universal R-matrix for quantum untwisted affine Lie algebras
      Functional Analysis and Its Applications 26, 69–71
      (1992). https://doi.org/10.1007/BF01077085

\bibitem{Hanany:1996ie}
A.~Hanany and E.~Witten,
%``Type IIB superstrings, BPS monopoles, and three-dimensional gauge dynamics,''
Nucl. Phys. B \textbf{492} (1997), 152-190
doi:10.1016/S0550-3213(97)00157-0
[arXiv:hep-th/9611230 [hep-th]].


\bibitem{Zenkevich:2017ylb}
A.~Nedelin, S.~Pasquetti and Y.~Zenkevich,
%``T[SU(N)] duality webs: mirror symmetry, spectral duality and gauge/CFT correspondences,''
JHEP \textbf{02} (2019), 176
doi:10.1007/JHEP02(2019)176
[arXiv:1712.08140 [hep-th]].\\
F.~Aprile, S.~Pasquetti and Y.~Zenkevich,
%``Flipping the head of $T[SU(N)]$: mirror symmetry, spectral duality and monopoles,''
JHEP \textbf{04} (2019), 138
doi:10.1007/JHEP04(2019)138
[arXiv:1812.08142 [hep-th]].

\bibitem{Fukuda:2020czf}
M.~Fukuda, Y.~Ohkubo and J.~Shiraishi,
%``Non-Stationary Ruijsenaars Functions for $\kappa=t^{-1/N}$ and Intertwining Operators of Ding-Iohara-Miki Algebra,''
SIGMA \textbf{16} (2020), 116
doi:10.3842/SIGMA.2020.116
[arXiv:2002.00243 [math.QA]].

\bibitem{Assel:2014awa}
B.~Assel,
%``Hanany-Witten effect and SL(2, $\mathbb{Z}$) dualities in matrix models,''
JHEP \textbf{10} (2014), 117
doi:10.1007/JHEP10(2014)117
[arXiv:1406.5194 [hep-th]].
      
\bibitem{Costello}
      K.~Costello,
%``Supersymmetric gauge theory and the Yangian,''
      [arXiv:1303.2632 [hep-th]].\\
      K.~Costello, E.~Witten and M.~Yamazaki,
%``Gauge Theory and Integrability, I,''
ICCM Not. \textbf{06} (2018) no.1, 46-119
doi:10.4310/ICCM.2018.v6.n1.a6
[arXiv:1709.09993 [hep-th]].\\
K.~Costello, E.~Witten and M.~Yamazaki,
%``Gauge Theory and Integrability, II,''
ICCM Not. \textbf{06} (2018) no.1, 120-146
doi:10.4310/ICCM.2018.v6.n1.a7
[arXiv:1802.01579 [hep-th]].\\
K.~Costello and M.~Yamazaki,
%``Gauge Theory And Integrability, III,''
[arXiv:1908.02289 [hep-th]].

      
    \bibitem{Bazhanov:2005as}
V.~V.~Bazhanov and S.~M.~Sergeev,
%``Zamolodchikov's tetrahedron equation and hidden structure of quantum groups,''
J. Phys. A \textbf{39} (2006), 3295-3310
doi:10.1088/0305-4470/39/13/009
[arXiv:hep-th/0509181 [hep-th]].

\bibitem{Gavrylenko:2020eov}
P.~Gavrylenko, M.~Semenyakin and Y.~Zenkevich,
%``Solution of tetrahedron equation and cluster algebras,''
[arXiv:2010.15871 [nlin.SI]].

\end{thebibliography}
\end{document}